\documentclass[11pt,tightenlines,eqsecnum,floats,aps,amsmath,amssymb,nofootinbib,prd,shownopacs,floatfix]{revtex4-2}

\usepackage{graphicx}
\usepackage{epstopdf}
\usepackage{latexsym}
\usepackage{amssymb}
\usepackage{amsmath}
\usepackage{color}
\usepackage{mathrsfs}
\usepackage{xparse}
\usepackage{float}
\usepackage{mathtools}
\usepackage{multirow}
\usepackage[center]{subfigure}

\usepackage{xcolor}
\begin{document}
\renewcommand\arraystretch{2}
 \newcommand{\bq}{\begin{equation}}
 \newcommand{\eq}{\end{equation}}
 \newcommand{\bqn}{\begin{eqnarray}}
 \newcommand{\eqn}{\end{eqnarray}}
 \newcommand{\nb}{\nonumber}
 \newcommand{\cb}{\color{blue}}
    \newcommand{\cc}{\color{cyan}}
     \newcommand{\lb}{\label}
        \newcommand{\cm}{\color{magenta}}
\newcommand{\rc}{\rho^{\scriptscriptstyle{\mathrm{I}}}_c}
\newcommand{\rd}{\rho^{\scriptscriptstyle{\mathrm{II}}}_c}
\NewDocumentCommand{\evalat}{sO{\big}mm}{%
  \IfBooleanTF{#1}
   {\mleft. #3 \mright|_{#4}}
   {#3#2|_{#4}}%
}
\newcommand{\PRL}{Phys. Rev. Lett.}
\newcommand{\PL}{Phys. Lett.}
\newcommand{\PR}{Phys. Rev.}
\newcommand{\CQG}{Class. Quantum Grav.}
\newcommand{\parallelsum}{\mathbin{\!/\mkern-5mu/\!}}
\title{Constraining regularization ambiguities in Loop Quantum Cosmology via CMB}
\author{Bao-Fei Li $^{1}$}
\email{libaofei@zjut.edu.cn}
\author{Meysam Motaharfar $^{2}$}
\email{mmotah4@lsu.edu}
\author{Parampreet Singh$^{2}$}
\email{psingh@lsu.edu}
\affiliation{
$^{1}$ Institute for Theoretical Physics $\&$ Cosmology, Zhejiang University of Technology, Hangzhou, 310023, China\\
$^{2}$ Department of Physics and Astronomy, Louisiana State University, Baton Rouge, LA 70803, USA}

\begin{abstract}

In order to investigate the potential observational signals of different regularization ambiguities in loop quantum cosmological models, we systematically compute and compare the primordial scalar power spectra and the resulting angular power spectra in the standard LQC and Thiemann regularized versions, modified LQC-I/II (mLQC-I/II), using both the dressed metric and the hybrid approaches. All three loop quantum cosmological models yield a non-singular bounce with a post-bounce physics that converges rapidly in a few Planck seconds. Using Starobinsky potential and the initial conditions for the background dynamics chosen to yield the same inflationary e-foldings, which are fixed to be $65$ in all three LQC models, we require that all three models result in the same scale-invariant regime for the primordial power spectrum with a relative difference of less than one percent. This permits us to explore the differences resulting from the deep Planck regime in the angular power spectrum. For the adiabatic states, our results demonstrate that the angular power spectrum predicted by the hybrid approach has a smaller deviation from the angular power spectrum predicted by the standard $\Lambda$CDM cosmological model at large angles in comparison with the dressed metric approach for all three models. The angular power spectrum predicted by mLQC-I in both the hybrid and the dressed metric approaches shows the smallest deviation from the one predicted by the standard $\Lambda$CDM cosmological model at large angular scales, except for the case of fourth order adiabatic initial states in the hybrid approach. On the contrary, mLQC-II results in the largest deviations for the amplitude of the angular power spectrum at large angles and is most disfavored.  

\end{abstract}

\maketitle

\section{Introduction}

Inflation, a finite quasi-de Sitter expanding phase, which, in its simplest form, is realized by a single scalar field known as the ``inflaton", evolving under the influence of a plateau-like potential, not only resolves longstanding puzzles in the standard big bang cosmology but also serves as a causal mechanism to seed the acoustic peaks in the Cosmic Microwave Background (CMB) and accounts for the distribution of Large-Scale Structure (LSS) from the evolution of primordial quantum vacuum fluctuations \cite{Guth:1980zm, Linde:1981mu, Bardeen:1983qw}. However, classical inflationary spacetimes are past-incomplete \cite{Borde:2001nh}, and the big bang singularity is inevitable when the universe evolves backward to the regime where the energy density and spacetime curvature diverge close to the Planck regime. To address this issue, one solution is to extend the inflationary spacetimes to the Planck regime by considering the quantum geometry effects. One of the most successful attempts to achieve this goal is Loop Quantum Cosmology (LQC), where the techniques of Loop Quantum Gravity (LQG) are applied to symmetry-reduced cosmological spacetimes \cite{Ashtekar:2006rx, Ashtekar:2006uz, Ashtekar:2006wn, Ashtekar:2011ni}. A key prediction of LQC is that the big bang singularity is replaced with a quantum bounce as spacetime curvature approaches the Planck regime, extending the spacetime to the contracting branch \cite{Ashtekar:2006wn, Ashtekar:2007em}, with the probability for the occurrence of the bounce turns out to be unity in the consistent histories formulation of quantum mechanics \cite{Craig:2013mga}. 

At the fundamental level, singularity resolution in LQC arises due to the underlying discreteness emerging from quantum geometry, whereby the evolutionary equation turns out to be  a second order discrete quantum difference equation. 
Interestingly, under reasonable conditions, the underlying quantum evolution can be accurately captured for a class of semi-classical states using effective dynamics \cite{Taveras:2008ke,Diener:2014mia,Diener:2014hba,Diener:2017lde,Singh:2018rwa}. In fact, extensive numerical simulations show that the effective spacetime description matches the GR trajectory to a great accuracy as soon as spacetime curvature becomes 1$\%$ of the Planck value \cite{Ashtekar:2006wn, Diener:2014mia}. This implies that the quantum geometric corrections to background dynamics quickly diminish away from the Planck regime. In addition, the effective dynamics accurately captures the underlying dynamics resulting from the quantum Hamiltonian constraint starting from the bounce point if one considers states which are sharply peaked at late times on the classical trajectory. This can be understood by comparing the expectation value of volume observable in the quantum theory and the bounce volume in effective dynamics. For states which are sharply peaked at late times, this difference is negligible \cite{Diener:2014hba}. In this manuscript, we assume the validity of effective dynamics in the entire regime which is true when one considers states which are sharply peaked at late times in the classical regime.

The phenomenological implications of LQC for various models have been extensively studied, assuming the validity of the effective dynamics \cite{Li:2021mop, Li:2023dwy}. The bounds on the energy density, expansion, and shear scalar in different models have been found \cite{Gupt:2011jh, Joe:2014tca, Singh:2013ava}, and strong curvature singularities have been shown to be generically resolved in isotropic models \cite{Singh:2009mz, Singh:2010qa, Saini:2018tto} as well as anisotropic models \cite{Singh:2011gp, Saini:2017ipg, Saini:2017ggt, Saini:2016vgo}. Moreover, it has been demonstrated that a viable non-singular inflationary model can be constructed in an isotropic model \cite{Singh:2006im, Ashtekar:2009mm, Ashtekar:2011rm, Giesel:2020raf} as well as Bianchi-I spacetime \cite{Gupt:2013swa, Linsefors:2014tna}. Effective dynamics also plays an important role in exploring quantum geometric effects in cosmological perturbations in LQC \cite{Agullo:2016tjh, Agullo:2023rqq} where different approaches exist \footnote{See also \cite{Tsujikawa:2003vr, Bojowald:2006tm, Magueijo:2007wf, Dapor:2013pka} as examples of other approaches.}: the deformed algebra approach \cite{Bojowald:2008gz, Cailleteau:2012fy, Cailleteau:2011kr}, the separate universe approach \cite{Wilson-Ewing:2016yan}, the hybrid approach \cite{Fernandez-Mendez:2013jqa, Gomar:2014faa, Gomar:2015oea, ElizagaNavascues:2018bgp}, and the dressed metric approach \cite{Agullo:2012sh, Agullo:2012fc, Agullo:2013ai}. However, among these approaches, the latter two are the most widely used to investigate the phenomenological predictions of LQC models which have been also used to address anomalies in CMB \cite{Ashtekar:2016wpi, Ashtekar:2020gec, Agullo:2020fbw, Agullo:2020cvg, Martin-Benito:2023nky}. 
Despite these successes, as any quantum theory, loop quantization of cosmological models faces the issue of quantization ambiguities. 
A class of these ambiguities in the background dynamics in standard LQC have been adequately addressed in isotropic \cite{Corichi:2008zb, Corichi:2009pp}, anisotropic models \cite{Motaharfar:2023hil}, as well as black hole spacetimes \cite{Li:2021snn}, however certain important ambiguities still remain. In this manuscript our focus is to understand potential observational imprints two of these: the regularization ambiguity arising from the treatment of Euclidean and Lorentzian terms in the Hamiltonian constraint, and the quantization ambiguities which lead to different effective mass functions in perturbations. 

Let us first discuss the regularization ambiguities. 
In the standard LQC model, the Lorentzian and Euclidean terms in the Hamiltonian constraint are combined using classical symmetry before quantization in Friedmann-Lemaître-Robertson-Walker (FLRW) spacetime \cite{Ashtekar:2003hd}. However, if two terms are treated independently during the quantization, they lead to different in-equivalent quantizations of LQC. Two notable examples are the so-called modified LQC-I (mLQC-I) and the modified LQC-II (mLQC-II) \cite{Yang:2009fp, Li:2018fco}, both arising from the Thiemann's regularization of the Hamiltonian constraint \cite{Thiemann:1996aw, Thiemann:1996av, Giesel:2006uj}. While classical identities on gravitational phase space are used to write the extrinsic curvature of the Lorentzian term of the Hamiltonian constraint in terms of holonomies in mLQC-I, the symmetry between extrinsic curvature and Ashtekar-Barbero connection in spatially flat spacetime is used to express the Lorentzian part in terms of holonomies in mLQC-II. In this sense, mLQC-I is closer to construction followed in full LQG and mLQC-II is only valid in spatially flat models. It has been demonstrated that the strong singularities are resolved \cite{Saini:2018tto} and the occurrence of inflation is generic in mLQC-I/II, like the standard LQC \cite{Li:2018fco}. In contrast to the second order discrete quantum difference equation in the standard LQC, the quantum Hamiltonian constraint in mLQC-I/II yields a fourth order discrete quantum difference equation \cite{Saini:2019tem}. Moreover, the effective modified Friedmann equation contains higher order terms of energy density for mLQC-I/II, while in the standard LQC only the quadratic term of the energy density appears \cite{Li:2018opr}. In addition, the maximum energy density at the quantum bounce in mLQC-I/II is different from the maximum energy density in the standard LQC. In contrast to the dynamics of mLQC-II, which shares qualitative similarities with standard LQC, mLQC-I exhibits notable differences. Particularly in the contracting phase, mLQC-I gives rise to an emergent quasi-de Sitter spacetime that arises with a Planckian value, implying that the contracting phase in mLQC-I is purely a quantum regime without a classical regime \cite{Assanioussi:2019iye}. While the nature of the bounce is asymmetric in mLQC-I \cite{Dapor:2017rwv}, the background dynamics is symmetric in the pre-bounce and post-bounce branches for mLQC-II as it is in the standard LQC. Considering such regularization ambiguities with different physical implications, the pertinent question is how the quantum effects of spacetime encoded in the pre-inflationary phase modify the dynamics of cosmological perturbations for each regularization. To answer this question, one needs to carefully examine how different modifications to the Hamiltonian constraint lead to modifications of the primordial power spectrum.

Apart from regularization ambiguities in the background Hamiltonian, there are also quantum ambiguities related to treatments of cosmological perturbations. Since LQC is a quantization of symmetry-reduced homogeneous spacetimes and the full connection to LQG is not established yet, in the two popular approaches -- dressed metric and hybrid approaches, linear perturbations are treated using Fock quantization in the loop quantized background.  The dressed metric approach is based on the Hamiltonian formulation of classical perturbations in the Arnowitt-Deser-Misner (ADM) phase space, in which the lapse and shift vector are treated as Lagrange multipliers. In this approach, the Hamiltonian constraint is expanded up to the second order in perturbations, then the zeroth order Hamiltonian constraint is loop quantized, while the second order constraint describes the dynamics of linear inhomogeneous perturbations \cite{Agullo:2012sh, Agullo:2012fc, Agullo:2013ai}. In other words, the inhomogeneous degrees of freedom can be interpreted as the quantum degrees of freedom propagating in the quantum spacetime described by the dressed metric after quantization. In fact, for sharply peaked states, the evolution of the scale factor in the dressed metric is governed by the effective dynamics in LQC. Although the hybrid and the dressed metric approaches share common features in the sense that the background geometry is loop quantized while the linear inhomogeneous perturbations are Fock quantized, there are some differences between these two approaches.  
In fact, in the hybrid approach, the background geometry is loop quantized, the zeroth mode of the scalar field is quantized in the standard Schr\"odinger representation, and the inhomogeneous perturbations are Fock quantized \cite{Fernandez-Mendez:2013jqa, Gomar:2014faa, Gomar:2015oea}. The solution to the resulting quantum dynamical equation is then solved by using the Born-Oppenheimer ansatz, which approximates the physical state as a direct product of the quantum background state and the states only depending on the gauge invariant modes. 
Whether or not the differences at the quantum level translate to any differences in predictions is a question which remains open, but at the practical level the difference in two approaches is tied to the way polymerization is performed in different steps to reach quantum geometry modifications to the Mukhanov-Sasaki equation \cite{Li:2022evi}. In the latter sense, signatures in CMB resulting from dressed and hybrid approaches can be seen as originating from quantization ambiguities which affect the effective mass functions in the Mukhanov-Sasaki equation \cite{Li:2022evi, Li:2023res}. 

The effects of regularization and quantum ambiguities on the primordial power spectrum were earlier studied in literature, and the primordial power spectrum in the standard LQC and mLQC-I/II has been computed for the dressed metric approach \cite{Li:2019qzr} (see also \cite{Agullo:2018wbf}) as well as the hybrid approach \cite{Li:2020mfi}. It was found that the resulting primordial power spectra in the standard LQC and mLQC-I/II have similar patterns with three distinctive regimes: the infrared regime, the intermediate oscillatory regime (the enhanced regime), and the ultraviolet regime (the scale-invariant regime where the predicted primordial power spectrum by LQC models is well approximated by a power law power spectrum). In fact, with the adiabatic initial states, both approaches predict an oscillating pattern of the primordial power spectrum with amplified amplitude in the regime preceding the observed scale-invariant primordial power spectrum in the CMB. Moreover, it has been shown that all LQC models predict the same amount of inflation and the scale-invariant regime in the primordial power spectrum in the ultraviolet regime, while the shape and amplitude of the primordial power spectrum are distinct for each regularization and quantum ambiguity in the infrared and the intermediate regimes. Since the comoving Hubble horizon is shrinking at the present time due to the accelerating expansion of the universe, these super-horizon modes with amplified amplitude can only be observed indirectly via non-Gaussianity effects \cite{Zhu:2017onp, Wu:2018sbr, Agullo:2017eyh}. Although it is not possible to directly detect such quantum gravitational effects in the primordial power spectrum via the CMB data, they can be used to constrain the regularization and quantum ambiguities in LQC models. In fact, it is expected that the modification of the infrared and the intermediate regimes of the primordial power spectrum leads to the modification of the angular power spectrum at large angles, i.e., low $l$ multipoles. Hence, one can compare the angular power spectrum predicted by different regularizations and quantum ambiguities in LQC models with the angular power spectrum predicted by the standard $\Lambda$CDM cosmological model to constrain regularization and quantum ambiguities. Therefore, in this study, we revisit the primordial power spectrum for LQC models in both the hybrid and the dressed metric approaches to understand the effects of different regularizations and quantum ambiguities, adiabatic initial states, and the initial time for which the adiabatic initial states are imposed in the contracting branch, in the infrared and the intermediate regimes of the primordial power spectrum. Then, by calculating the relevant angular power spectrum, we aim to understand for which model the predicted angular power spectrum has more or less  compatibility with the angular power spectrum predicted by the standard $\Lambda$CDM cosmological model at large angular scales. 

To understand quantization ambiguities, we calculate the effective mass function for each model in both the hybrid and the dressed metric approaches for the Starobinsky potential while solving the background dynamics by fixing the initial conditions at the bounce. We then compute the scalar primordial power spectrum \footnote{Since we only compute the primordial power spectrum for scalar perturbations in this manuscript, henceforth, when we refer to the ``primordial power spectrum," it specifically refers to the scalar primordial power spectrum.} by imposing zeroth, second, and fourth order adiabatic initial states in the contracting branch with an exception to the mLQC-I in the dressed metric approach, whose initial state is chosen to be the exact de Sitter solution tailored to the special properties of the effective mass function in this model and approach. We compare the primordial power spectrum predicted by these three models by tuning the inflaton's mass and the initial value of the inflaton field at the bounce in such a way that all models predict the same number of e-foldings, $N_{e} = 65$, and also the same scale-invariant regime with the relative difference in the power less than one percent in that regime. In this sense, we can compare the primordial power spectrum for each regularization and quantum ambiguity in the infrared and the intermediate regimes, because of which any modifications in the angular power spectrum can be merely due to such modifications in the infrared and the intermediate regimes. In doing so, we find that the shape and amplitude of the primordial power spectrum in the infrared and the intermediate regimes depend on the regularization and quantum ambiguities, the order of adiabatic initial states, and how far from the bounce they are imposed in the contracting branch. In fact, it is found that the intermediate regime in the primordial power spectrum has a larger amplitude if the initial states are imposed further away from the bounce in the contracting branch, except for mLQC-I in the dressed metric approach. In addition, we find that the intermediate regime has a smaller amplitude for higher order adiabatic initial states in both LQC and mLQC-II and for both the hybrid and the dressed metric approaches. Moreover, we realize that the primordial power spectrum has a spike preceding the scale-invariant regime in the case of fourth order adiabatic initial states for mLQC-I model and the hybrid approach. Furthermore, it turns out  that the primordial power spectrum for both LQC and mLQC-II in the dressed metric approach has a slightly stronger suppressing regime in the infrared regime in comparison with the hybrid approach. Finally, for mLQC-I in the dressed metric approach with de Sitter initial state, the primordial power spectrum reaches a constant value in the infrared regime rather than being suppressed. 

Motivated by the fact that each model has different predictions for the primordial power spectrum in the infrared and the intermediate regimes, we aim at finding the relevant angular power spectrum to investigate which model has most compatibility with the angular power spectrum predicted by the standard $\Lambda$CDM cosmological model. Hence, we feed the calculated primordial power spectrum into the CAMB code  \footnote{Code for Anisotropies in the Microwave Background (CAMB) is used to calculate cosmological quantities by solving background and perturbation equations. For details see, https://camb.readthedocs.io.} as an external power spectrum and compute the angular power spectrum. We observe larger amplitude at large angles, i.e., low $l$ multipoles, for all models in both the hybrid and dressed metric approaches in comparison with the angular power spectrum predicted by the standard $\Lambda$CDM cosmological model. However, our results demonstrate that the angular power spectrum predicted by the hybrid approach has a smaller deviation from the angular power spectrum predicted by the standard $\Lambda$CDM cosmological model at large angles in comparison with the dressed metric approach for all three models, except for the case of mLQC-I with fourth order adiabatic initial states in the hybrid approach. Moreover, among these three models, the angular power spectrum predicted by mLQC-I in both the hybrid and the dressed metric approaches shows the smallest deviation from the angular power spectrum predicted by the standard $\Lambda$CDM cosmological model at large angles, except for the case with fourth order adiabatic initial states in the hybrid approach due to the presence of a spike preceding the scale-invariant regime. On the contrary, based on our results, mLQC-II is disfavored by the data since it predicts the largest amplitude for the angular power spectrum at large angles among these three models. Therefore, we conclude that the regularization used in mLQC-I is preferred from an observational perspective since the predicted angular power spectrum has the smallest deviation at large angles from the angular power spectrum predicted by the standard $\Lambda$CDM cosmological model.

The manuscript is organized as follows. In Sec. \ref{section II}, we briefly review the effective dynamics of background spacetime for three LQC models arising from regularization ambiguities, namely the standard LQC, mLQC-I, and mLQC-II. In Sec. \ref{sectionIII}, we briefly review the cosmological perturbations in both the hybrid and the dressed metric approaches and present the effective mass function in the Mukhnov-Sasaki equation for each model and approach. In Sec. \ref{sec:power_spectrum}, we numerically solve the background dynamics and the Mukhanov-Sasaki equation of the linear perturbations for different initial states set in the contracting phase to compute the primordial power spectrum for each model and approach. Then, we compute the relevant angular power spectrum by feeding the numerical primordial power spectrum into the CAMB code. In this way, we show how different regularizations and quantum ambiguities lead to different predictions for the angular power spectrum at large angles and compare the results with the angular power spectrum predicted by the standard $\Lambda$CDM cosmological model. Finally, we give a summary and conclusion in Sec. \ref{sec:conclusion}. In this manuscript, we use the Planck units with $\hbar=c=1$ while keeping Newton's constant $G$ explicitly. 

\section{A Brief Review of Loop Quantum Cosmology in \\ Different Regularizations} 
\lb{section II}

In this section, we briefly review the effective dynamics of the background spacetime in  three distinct loop quantum cosmological models for a spatially flat, homogeneous, and isotropic FLRW spacetime: the standard LQC and the modified LQCs (mLQCs), namely mLQC-I and mLQC-II. These models originate from different regularizations of the classical Hamiltonian constraint in LQG for FLRW spacetime. As is well known, the classical Hamiltonian constraint in LQG is composed of two parts, namely the Euclidean term and the Lorentzian term. In the standard LQC, these two terms are combined using classical symmetry and then loop quantized, while in mLQCs, separate regularizations of the Lorentzian term are implemented, which result in two distinct variants of the loop quantum cosmological model, namely mLQCs \cite{Yang:2009fp}. For these three models, the evolution of quantum dynamics is governed by their own discrete quantum difference equations. Moreover, when the background state is chosen to be the semi-classical state, the main properties of the quantum evolution of loop quantum cosmological models can be faithfully captured by the effective dynamics governed by an effective Hamiltonian constraint, which can be obtained from the polymerization of the classical Hamiltonian constraint of the FLRW spacetime. This effective description provides a convenient approach to studying the phenomenological implications of the loop quantum cosmological models, as they have been frequently used in the literature (see, e.g., \cite{Li:2018opr,Li:2018fco,Li:2019ipm}). In the following, we briefly review the dynamical equations of the effective dynamics in the standard LQC and mLQCs.

\subsection{The effective dynamical equations in LQC}

For a spatially flat, homogeneous, and isotropic FLRW universe filled with a massive scalar field, the classical phase space is spanned by four degrees of freedom, which can be chosen as $\{v,b,\phi,p_\phi\}$, where $v$ denotes the volume of the universe (related to the scale factor $a$, i.e., $v = a^3$) with its conjugate momentum $b$ which equals $\gamma H$ in the classical theory with $\gamma$ being the Barbero-Immirzi parameter and $H =\dot a/a$ the Hubble rate. Besides, $\phi$ stands for the scalar field, and $p_\phi$ is its conjugate momentum. These canonical variables satisfy the standard Poisson brackets $\{b,v\}=4\pi G\gamma$ and $\{\phi,p_\phi\}=1$. The classical dynamics of a spatially flat FLRW universe filled with a single scalar field is based on the classical Hamiltonian constraint that takes the form
\bq
\lb{classical_Hamiltonian}
\mathcal H=-\frac{3 v b^2}{8 \pi G \gamma^2 } + \frac{p^2_{\phi}}{2 v} + v U(\phi),
\eq
where $U(\phi)$ represents the potential of the scalar field. From this Hamiltonian, it is straightforward to obtain the corresponding Hamilton's equations and the classical Friedmann equation. Then the effective Hamiltonian constraint in the standard LQC can be formally obtained from the polymerization of the momentum $b$ in the classical Hamiltonian constraint. To be specific, the   rule of thumb to obtain the effective Hamiltonian constraint in the standard LQC is to apply the polymerization ansatz $b^2\rightarrow \sin^2(\lambda b)/\lambda^2$ in the classical Hamiltonian constraint (\ref{classical_Hamiltonian}). In this way, one can recover the effective Hamiltonian constraint of standard LQC in the $\bar \mu$ scheme \cite{Ashtekar:2006wn}, namely
\bq
\lb{LQC_Hamiltonian}
\mathcal{H}_\mathrm{LQC}= -\frac{3 v \sin^2 (\lambda b)}{8 \pi G \gamma^2 \lambda^2} + \frac{p^2_{\phi}}{2 v} + v U(\phi).
\eq
Correspondingly, it is straightforward to obtain the Hamilton's equations in LQC, which turn out to be
\bqn
\dot v &= &\frac{3v}{2 \gamma \lambda} \sin (2 \lambda b),\quad \quad
\dot b = - 4 \pi G \gamma \left(\rho+P\right),\\
\dot \phi &=& \frac{p_{\phi}}{v},~~~~~~~~~~~~~~~~~
\dot p_{\phi} = -v U_{,\phi},
\eqn
where $U_{,\phi}$ stands for the derivative of the potential with respect to the scalar field. Besides, in the  Hamilton's equations, the energy density and the pressure of the scalar field are given, respectively, by
\bq
\lb{energy_pressure}
\rho=\frac{p^2_{\phi}}{2 v^2} + U(\phi),~~~P=\frac{p^2_{\phi}}{2 v^2} -U(\phi).
\eq
From the Hamilton's equation for volume and the relation between the energy density and the momentum $b$, one can obtain the modified Friedmann equation in LQC, which takes the shape
\bq
\lb{Friedmann_LQC}
H^2 = \frac{8\pi G}{3}\rho\left(1- \frac{\rho}{\rho_{c}}\right),
\eq
here $\rho_c=3/8\pi G \lambda^2 \gamma^2\approx 0.41 \rho_{\mathrm{Pl}}$ is the maximum energy density at which the quantum bounce takes place in LQC. The modified Friedmann equation is well suited for studying the general aspects of the inflationary scenario in LQC, and one only needs to make use of  different types of the inflationary potentials to investigate the extension of the relevant inflationary spacetimes to the Planck regime. For the actual numerical simulation of the inflationary universe in LQC, the initial conditions are usually set right at the bounce point, where the parameter space is essentially a one-parameter space spanned by the value of the scalar field. More details on setting up the initial conditions for the background evolution will be discussed in Sec. \ref{sec:power_spectrum}.
 
\subsection{The effective dynamical equations in mLQCs}

The effective dynamical equations for mLQC-I and mLQC-II have been obtained in Ref. \cite{Li:2018opr, Li:2018fco}. In particular, for mLQC-I, the effective Hamiltonian constraint takes the form 
\bq
\mathcal{H}_{\scriptscriptstyle{\mathrm{I}}} = \frac{3 v}{8 \pi G \lambda^2} \left\{\sin^2(\lambda b) -  \frac{(\gamma^2+1) \sin^2(2 \lambda b)}{4\gamma^2}\right\} + \frac{p^2_{\phi}}{2 v} + v U(\phi).
\eq
Since the matter sector remains unaltered as in the standard LQC, the above modified Hamiltonian constraint in mLQC-I only changes the dynamical equations in the geometric sector. It is straightforward to check that the resulting Hamilton's equations are given explicitly by
\bq
\dot v = \frac{3 v \sin(2 \lambda b)}{2 \gamma \lambda } \Big\{ (\gamma^2+1) \cos(2\lambda b) - \gamma^2\Big\},\quad 
\dot b= - 4 \pi G\gamma \left(\rho+ P\right),
\eq 
where $\rho$ and $P$ are still given by their standard definitions in (\ref{energy_pressure}). Although the unique properties of the background dynamics of the mLQC-I model already become manifest in the numerical simulations using Hamilton's equations, one can reach a deeper understanding of its dynamical features only when its modified Friedmann equation becomes available. It turns out that, in contrast to the standard LQC, where both the contracting and the expanding branches are described by the same modified Friedmann equation (\ref{Friedmann_LQC}), these two branches are actually governed by different modified Friedmann equations in mLQC-I. In particular, the modified Friedmann equation in the expanding (post-bounce) branch is given by \cite{Li:2018opr}
\bq
H^2_\mathrm{post} = \frac{8\pi G}{3} \rho \left(1- \frac{\rho}{\rho_{c}^{\scriptscriptstyle{\mathrm{I}}}}\right) \left [1 + \frac{\gamma^2 \rho/\rho_{c}^{\scriptscriptstyle{\scriptscriptstyle{\mathrm{I}}}}}{(1+ \gamma^2) (1+ \sqrt{1- \rho/\rho_{c}^{\scriptscriptstyle{\scriptscriptstyle{\mathrm{I}}}})}^2}\right ],
\eq
where $\rho_{c}^{\scriptscriptstyle{\mathrm{I}}}=\rho_c/4(\gamma^2+1)$ is the maximum energy density at which the quantum bounce takes place in mLQC-I. The above modified Friedmann equation reduces to the classical Friedmann equation when the energy density is far below the Planck energy density. Whereas in the contracting (pre-bounce) phase the corresponding modified Friedmann equation takes the form 
\bq
H^2_\mathrm{pre} = \frac{8\pi G \alpha \rho_{\Lambda}}{3} \left(1- \frac{\rho}{\rho_{c}^{\scriptscriptstyle{\mathrm{I}}}}\right) \left [1 + \frac{\rho (1 - 2 \gamma^2 + \sqrt{1- \rho/\rho_{c}^{\scriptscriptstyle{\mathrm{I}}}})}{4 \gamma^2 \rho_{c}^{\scriptscriptstyle{\mathrm{I}}} (1 + \sqrt{1- \rho/\rho_{c}^{\scriptscriptstyle{\mathrm{I}}}}) } \right],
\eq
with $\alpha = \frac{1-5\gamma^2}{1+\gamma^2}$ and $\rho_{\Lambda} = \frac{3}{8 \pi G \alpha \lambda^2 (1+\gamma^2)^2}$. This implies an emergence of the effective cosmological constant $\rho_{\Lambda}$ and a rescaled Newton's constant $\tilde G=\alpha G$ in the asymptotic region of the contracting phase when $\rho\ll\rho_{c}^{\scriptscriptstyle{\mathrm{I}}}$. The different asymptotic behavior of the modified Friedmann equation in the expanding and the contracting branches in mLQC-I provides an intuitive explanation for the asymmetric evolution of the mLQC-I universe with respect to the quantum bounce. Although the classical universe is recovered in the future of the expanding branch, there exists only a quasi-de Sitter phase in the past of the contracting branch when the universe is filled with matter satisfying the weak energy condition. 

On the other hand, the effective Hamiltonian for mLQC-II can be written as follows 
\bq
\mathcal{H}_{\scriptscriptstyle{\mathrm{II}}} = -\frac{3 v}{2 \pi G \gamma^2 \lambda^2} \sin^2\left(\frac{\lambda b}{2}\right) \left\{ 1+ \gamma^2 \sin^2\left(\frac{\lambda b}{2}\right)\right\} + \frac{p^2_{\phi}}{2 v} + v U(\phi).
\eq
Correspondingly, the Hamilton's equations in the geometric sector are given by
\bq
\dot v = \frac{3 v \sin( \lambda b)}{\gamma \lambda}\left\{1+ \gamma^2 - \gamma^2 \cos(\lambda b)\right\},~~~~
\dot b = - 4 \pi G \gamma \left(\rho+P\right).
\eq
Similar to the standard LQC, the modified Friedmann equation in mLQC-II in both contracting and the expanding branches takes the same form and is given by
\bq
H^2 = \frac{16\pi G}{3} \rho \left(1- \frac{\rho}{\rho_{c}^{\scriptscriptstyle{\mathrm{II}}}}\right) \left(\frac{1+ 4 \gamma^2 (1+ \gamma^2) \rho/\rho_{c}^{\scriptscriptstyle{\mathrm{II}}}}{1+ 2 \gamma^2 \rho/\rho_{c}^{\scriptscriptstyle{\mathrm{II}}} + \sqrt{1+ 4 \gamma^2 (1+ \gamma^2) \rho/\rho_{c}^{\scriptscriptstyle{\mathrm{II}}}}}\right),
\eq
with the maximum energy density $\rho_{c}^{\scriptscriptstyle{\mathrm{II}}} = 4(1+\gamma^2) \rho_{c}$. As a result, the background evolution of the mLQC-II universe is also symmetric with respect to the quantum bounce when the universe is coupled with a massless scalar field. Moreover, the previous studies have shown that the qualitative dynamics in two models are also very similar to each other as well \cite{Li:2019ipm}. Consequently, in order to distinguish mLQC-II from LQC, further information on the linear perturbations in the two theories is vital to revealing the quantitative difference in the predictions of these two models on the CMB observations. In particular, it is worthwhile to compare the predictions on the primordial power spectrum from all three models, namely LQC and mLQCs, which compose the main content of the next two sections. 

\section{The linear perturbation theories in loop quantum cosmological models: dressed metric approach vs hybrid approach} 
\lb{sectionIII}

In this section, we briefly review the Mukhanov-Sasaki equation for the linear cosmological perturbations in the dressed metric and the hybrid approaches in LQC and mLQCs. For a detailed exposition of the linear perturbation theory in these three models, we refer the readers to our previous work \cite{Li:2019qzr,Li:2020mfi, Li:2022evi, Li:2023res}. In the following, we only cite the main results that are relevant for numerically computing the primordial power spectrum and the relevant angular power spectrum in these models. It turns out that in all three models, the linear cosmological perturbations satisfy the modified Mukhanov-Sasaki equation, which is characterized by different effective mass functions that essentially originate from regularization ambiguities in the background dynamics and quantum ambiguities related to treatments of cosmological perturbations. In terms of the rescaled Mukhanov-Sasaki variable $\nu_k$ which is related to the comoving curvature perturbation $\mathcal R_{ k}$ via $\nu_{k}=z_s\mathcal R_{ k}$ with $z_s=a\dot{ \phi}/H$, the modified Mukhanov-Sasaki equation in each model takes the generic form
\bq
\lb{MS}
\nu^{\prime \prime}_{k} + (k^2 + s) \nu_{k} = 0,
\eq
where $s$ stands for the effective mass term, whose explicit form is both model and approach dependent. A prime in the above equation denotes differentiation with respect to the conformal time $\eta$. Moreover, the mode function is normalized according to the Wronskian condition
\begin{align}
\nu_{k}(\nu^{\prime}_{k})^{\star} - (\nu_{k})^{\star} \nu_{k}^{\prime} = i,
\end{align}
with the asterisk standing for the complex conjugate. In the actual simulations, we set the initial states in the contracting phase when the relevant modes of interest are inside the comoving Hubble horizon. Since the adiabatic condition is well satisfied for those modes, the initial states can be chosen as the adiabatic states, which are essentially the WKB solutions of Eq. (\ref{MS}), namely
\begin{align}
\lb{nu-solution}
\nu_{k} = \frac{1}{\sqrt{2 W_{k}}} e^{-i \int^{\eta} W_{k} (\bar \eta) d \bar \eta}.
\end{align}
Once plugging the above solution back into Eq. (\ref{MS}), one can obtain a differential equation of $W_{k}$ that takes the form
\begin{align}\lb{wk}
W_{k}^2 = k^2+ s - \frac{1}{2} \frac{W_{k}^{\prime\prime}}{W_{k}} + \frac{3}{4} \left(\frac{W^{\prime}_{k}}{W_{k}}\right)^2.
\end{align}

Starting from the zeroth order solution, $W_{k}^{(0)}=k$, the adiabatic solutions at the second and fourth orders can be obtained as
\begin{align}
W_{k}^{(2)} = \sqrt{k^2 + s},~~~~~
W_{k}^{(4)} = \frac{\sqrt{f(s, k)}}{4 |k^2+ s|},
\end{align}
where $f(s, k) = 5 {s^{\prime}}^2 + 16 k^4 (k^2 + 3s) + 16 s^2 (3k^2+ s) - 4 s^{\prime \prime} (k^2 + s) $. However, we should point out that the initial states of the perturbations in mLQC-I are chosen in the contracting branch, where the de Sitter phase is a very good approximation and the effective mass function is well approximated by $s = - 2/{\eta^2}$. Therefore, Eq. (\ref{MS}), has the exact solutions, which are \cite{Li:2019qzr}
\begin{align}
\lb{de-sitter-solution}
\nu_k=\alpha_k\frac{e^{-ik\eta}}{\sqrt{2k}}\left(1-\frac{i}{k\eta}\right)+\beta_k\frac{e^{ik\eta}}{\sqrt{2k}}\left(1+\frac{i}{k\eta}\right),
\end{align}
where $\alpha_{k}$ and $\beta_{k}$ are two integration constants. In our simulations, the initial states of the perturbations are chosen as the positive frequency modes with $\alpha_{k}=1$ and $\beta_{k}=0$.

Given the initial states in the contracting phase, one should propagate the modified Mukhanov-Sasaki equation until the end of inflation, where the power spectrum, i.e., the correlation function between two modes $P_{\nu_{k}}$, is evaluated for the observable modes that have re-entered the Hubble horizon at present. In order to compare the results to the observational data, it is common to use the power spectrum of the comoving curvature perturbation whose magnitude freezes for the super-horizon modes, and it can be computed from $P_{\nu_{k}}$ as
\bq
P_{{\mathcal{R}}_{k}} = \frac{P_{\nu_{k}}}{z_{s}^2} = \frac{k^3}{2\pi^2} \frac{|\nu_{k}|^2}{z_{s}^2}.
\eq
It should be noted that the above power spectrum is only valid in the regime where the adiabatic initial states are real numbers at the initial time, which means $k^2 + s\ge 0$ for $W_{k}^{(2)}$ and $f(s, k)\ge 0$ for $W_{k}^{(4)}$. Before we proceed with the numerical results of the power spectrum, we need to fix the effective mass term in each model in both the dressed metric and the hybrid approaches. Since these effective mass terms have already been discussed in detail in our previous work \cite{Li:2019qzr,Li:2020mfi, Li:2022evi, Li:2023res}, we briefly mention their explicit forms in both  approaches for three different LQC models in the following subsection.

\subsection{The dressed metric approach in loop quantum cosmological models: \\polymerization aspects}

In the dressed metric approach, the quantum fluctuations propagating on a quantum background spacetime can be equivalently described as propagating on a continuum spacetime with a dressed metric derived from effective dynamics. The general formalism is based on the Hamiltonian formulation of the perturbation theory in general relativity introduced by Langlois \cite{Langlois:1994ec}. Considering a single scalar field minimally coupled to gravity on a spatially flat FLRW background, the mass function of the classical Mukhanov-Sasaki equation is given by
\begin{align}
\lb{dressed_classical}
s = {\mathfrak U}^2 - \frac{a^{\prime\prime}}{a}, %= \Omega^2 - a \ddot a - \dot a^2,
\end{align}
where the term related to the potential of the scalar field reads
\bq
\lb{effective_potential}
{\mathfrak U}^2 = \frac{24 \pi G p_{\phi}^2}{a^4} - \frac{18 p_{\phi}^4}{a^6} \frac{1}{\pi_{a}^2} - 12 a p_{\phi} U_{,\phi} \frac{1}{\pi_{a}} + a^2 U_{,\phi\phi},
\eq
with $U_{,\phi}$ denoting the derivative of the potential with respect to the scalar field $\phi$.
The quantum ambiguities in the dressed metric approach are due to the presence of the inverse of the conjugate momentum of the scale factor. In fact, in classical theory, $\pi_a$ is directly related to $b$ via $\pi_a=-6a^2b/\kappa\gamma$. Correspondingly, $\pi^2_a$ is proportional to $b^2$. As already discussed in the last section, in the loop quantum cosmological models, the effective background Hamiltonian can be obtained from the polymerization of $b^2$ in the classical background Hamiltonian. Since the linear perturbations are propagating on the effective spacetimes as long as the effective dynamics is valid, the relevant background quantities in the classical Mukhanov-Sasaki equation must be polymerized in a manner consistent with the polymerization of the background dynamics. Therefore, in the classical effective potential (\ref{effective_potential}), $1/\pi^2_a$ must be polymerized in the same manner as in the background dynamics. In contrast, there is no information on a proper polymerization of $1/\pi_a$ from the background dynamics and thus it introduces quantum ambiguities into the effective mass function in the modified Mukhanov-Sasaki equation. In the following, we adopt the polymerization ansatz that was employed in our previous work \cite{Li:2019qzr}. To be specific, $1/\pi_a$ and $1/\pi^2_a$ are polymerized in LQC according to the ansatz
\bq
\frac{1}{\pi_{a}} \rightarrow -\frac{4\pi \gamma \lambda \cos(\lambda b)}{3 a^2 \sin(\lambda b)},~~~~~
\frac{1}{\pi_{a}^2} \rightarrow \frac{16 \pi^2 G^2 \gamma^2 \lambda^2}{9 a^4 \sin^2(\lambda b)},
\eq
and similarly for mLQC-I, we use the following ansatz
\begin{align}
\frac{1}{\pi_{a}^{\scriptscriptstyle{\mathrm{I}}}} &= \frac{8 \pi G \gamma \lambda \tilde\Theta(b)}{3a^2 \sqrt{(1+\gamma^2)\sin^2(2\lambda b) - 4 \gamma^2 \sin^2(\lambda b)}},\\
\frac{1}{{\pi_{a}^{\scriptscriptstyle{\mathrm{I}}}}^{2}} &= \frac{64 \pi^2 G^2 \gamma^2 \lambda^2}{9 a^4\left((1+\gamma^2) \sin^2(2\lambda b) - 4 \gamma^2 \sin^2(\lambda b)\right)},
\end{align}
with $\tilde\Theta(b) = 1 - 2(1+\gamma^2) \sin^2(\lambda b)$.
And finally, in the case of mLQC-II, we employ the ansatz
\begin{align}
\frac{1}{\pi_{a}^{\scriptscriptstyle{\mathrm{II}}}} &= -\frac{2 \pi \gamma \lambda \cos(\lambda b/2)}{3 a^2 \sin(\lambda b/2) \sqrt{1+\gamma^2 \sin^2(\lambda b/2)}},\\
\frac{1}{{\pi_{a}^{\scriptscriptstyle{\mathrm{II}}}}^2} &= \frac{4 \pi^2 \gamma^2 \lambda^2}{9 a^4 \sin^2(\lambda b/2) (1+ \gamma^2 \sin^2(\lambda b/2))}.
\end{align}
Once plugging the above polymerization ansatz into the classical mass function (\ref{dressed_classical}), one can obtain the corresponding effective mass function for each model in the dressed metric approach. Then, given an appropriate initial state in the contracting phase, we can numerically compute the evolution of the mode function throughout the evolution of the universe until the end of inflation.

\subsection{The Hybrid approach in loop quantum cosmological models: \\polymerization aspects}

The modified Mukhanov-Sasaki equation in the hybrid approach can be obtained by the following polymerization procedures, which are similar to those used in the dressed metric approach. After all, both approaches apply the same hybrid quantization of the background dynamics and the perturbations, i.e., the background is loop quantized in the $\bar \mu$ scheme while the perturbations are Fock quantized. The main distinctions between these two approaches at the level of effective dynamics originate from the different forms of the classical mass function, which are equivalent on the classical trajectories but lead to different effective mass functions after polymerization due to quantization ambiguities \cite{Li:2022evi}. Here we cite the expressions of the effective mass functions in the hybrid approach for LQC and mLQCs. The details of the derivation of the effective mass functions can be found in our previous work \cite{Li:2020mfi}. These effective mass functions are based on the polymerization of the classical mass function, which is cast into the form
\begin{equation}
\lb{classical_hybrid}
s = \frac{4 \pi G p_{\phi}^2}{3 v^{4/3}} \left(19 - 24 \pi G \gamma^2 \frac{p_{\phi}^2}{\Omega^2}\right) + v^{2/3} \left(U_{,\phi\phi} + \frac{16 \pi G \gamma p_{\phi} \Lambda}{\Omega^2} U_{,\phi} - \frac{16\pi G}{3} U\right),
\end{equation}
where $\Omega$ and $\Lambda$ are equal in the classical theory, and both of them are given by $vb$. We distinguish these two terms only for the convenience of the effective theory in which they are polymerized in different ways according to the model under consideration. To be specific, in the case of standard LQC, $\Omega$ and $\Lambda$ are polymerized to be
\bq
\Omega_\mathrm{LQC} = v \frac{\sin(\lambda b)}{\lambda},\quad \quad
\Lambda_\mathrm{LQC} = v \frac{\sin(2\lambda b)}{2\lambda}.
\eq
while in mLQC-I, they are polymerized into
\bq
\Omega^2_{\scriptscriptstyle{\mathrm{I}}} = - \frac{v^2 \gamma^2}{\lambda^2} \left\{\sin^2(\lambda b) - \frac{\gamma^2 +1}{4 \gamma^2} \sin^2(2\lambda b)\right\},\quad \quad
\Lambda_{\scriptscriptstyle{\mathrm{I}}} = v \frac{\sin(2\lambda b)}{2\lambda}.
\eq
Finally, in mLQC-II, they are polymerized as 
\bq
\Omega^2_{\scriptscriptstyle{\mathrm{II}}} = \frac{4v^2}{\lambda^2} \sin^2\left(\frac{\lambda b}{2}\right) \left\{ 1+ \gamma^2 \sin^2\left(\frac{\lambda b}{2}\right)\right\} ,\quad \quad
\Lambda_{\scriptscriptstyle{\mathrm{II}}} = v \frac{\sin(\lambda b)}{\lambda}.
\eq
Substituting the above polymerization ansatz for $\Omega$ and $\Lambda$ into the classical mass function (\ref{classical_hybrid}), one can obtain the effective mass function for each model.
Given the functionality for the effective mass term for each regularization, one is able to proceed to find the curvature power spectrum. Therefore, the goal of the next section is to compute the curvature power spectrum for each regularization in both the dressed metric and the hybrid approaches and then feed it into the CAMB code to calculate the relevant angular power spectrum.

\section{The primordial scalar power spectrum and the angular power spectrum in loop quantum cosmological models}
\lb{sec:power_spectrum} 

In this section, we study the primordial power spectrum and the relevant angular power spectrum of the loop quantum cosmological models discussed in the previous section. In Sec. \ref{sectionII-A}, we first numerically solve the background equations with the initial conditions set at the bounce, then solve the Mukhanov-Sasaki equation given in Eq. (\ref{MS}) with different effective mass functions and appropriate initial states for each model and approach. Finally, we compute the primordial scalar power spectrum and  the relevant angular power spectrum for each model and approach and compare the latter with the angular power spectrum predicted by the standard $\Lambda$CDM cosmological model in Sec. \ref{sectionIV-B}.

\subsection{The primordial power spectra in loop quantum cosmological models}\label{sectionII-A}

After reviewing the effective background dynamics and the modified Mukhanov-Sasaki equations in the standard LQC and mLQCs, we proceed to compare the observable predictions of these three models in both the hybrid and the dressed metric approaches in this section. In fact, we compute the primordial power spectrum and the relevant angular power spectrum for different regularizations and quantum ambiguities, along with appropriate initial states. As discussed earlier, the main differences among these models come from different quantizations of the gravitational sector as well as different polymerizations of the conjugate momentum of the scale factor in the Mukhanov-Sasaki equation. Such differences are encoded in the time-dependent effective mass function in the Mukhanov-Sasaki equation, which takes distinct forms for different regularizations and quantum ambiguities. Besides, to compute the primordial power spectrum, one first needs to fix the background dynamics for each model. Hence, we consider the extension of the inflationary scenario in the loop quantum cosmological models, with the inflationary phase driven by a single scalar field and the potential given by the Starobinsky potential 
\bq
U(\phi) = \frac{m^2}{32 \pi G} \left(1- e^{- \sqrt{\frac{16 \pi G}{3}}\phi}\right)^2.
\eq
This potential has only one free parameter, i.e., the inflaton's mass $m$, which will be determined by a phenomenological matching of the predicted  primordial power spectrum with the observational data. In fact, with the scalar power spectrum $A_{s}$ and scalar spectral index $n_{s}$ given at the pivot mode $k_{\star}/a_0 = 0.05~\mathrm{Mpc^{-1}}$ (here $a_0$ stands for the scale factor at present), respectively by \cite{Planck:2018jri}
\begin{align}\label{Planck}
\ln(10^{10}A_{s}) = 3.044\pm 0.014 \ (68\% CL), \ \ \ \ \ \ n_{s} = 0.9649 \pm 0.0042 \ (68\% CL),
\end{align}
according to the Planck collaboration in the TT,TE,EE-lowE+lensing 68\% limits data, one can fix the inflaton's mass to be $m = 2.44\times 10^{-6}$ (in Planck units) by simply matching analytical expressions for power spectrum and spectral index obtained using the slow-roll approximation with the observational values in Eq. (\ref{Planck}).

\begin{figure}
    \centering
    \includegraphics[scale = 0.4]{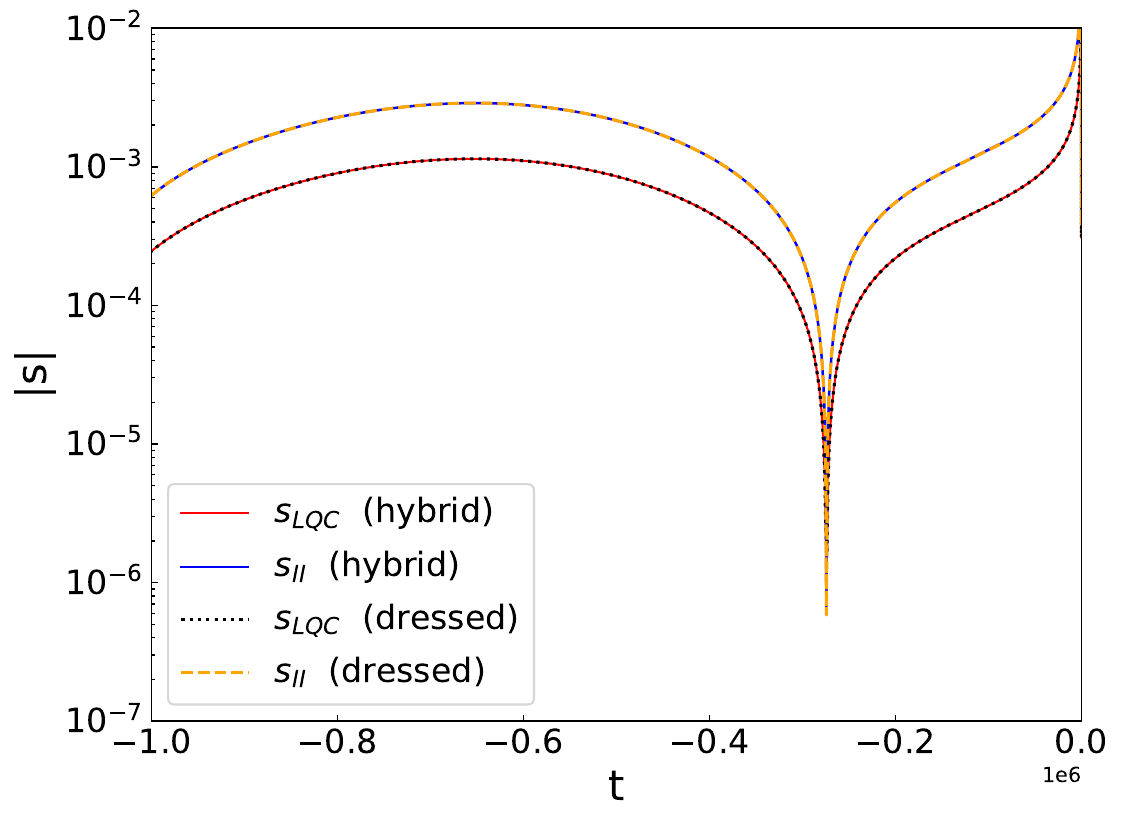} \ \ \ \ \ \ \ \ \ \includegraphics[scale = 0.4]{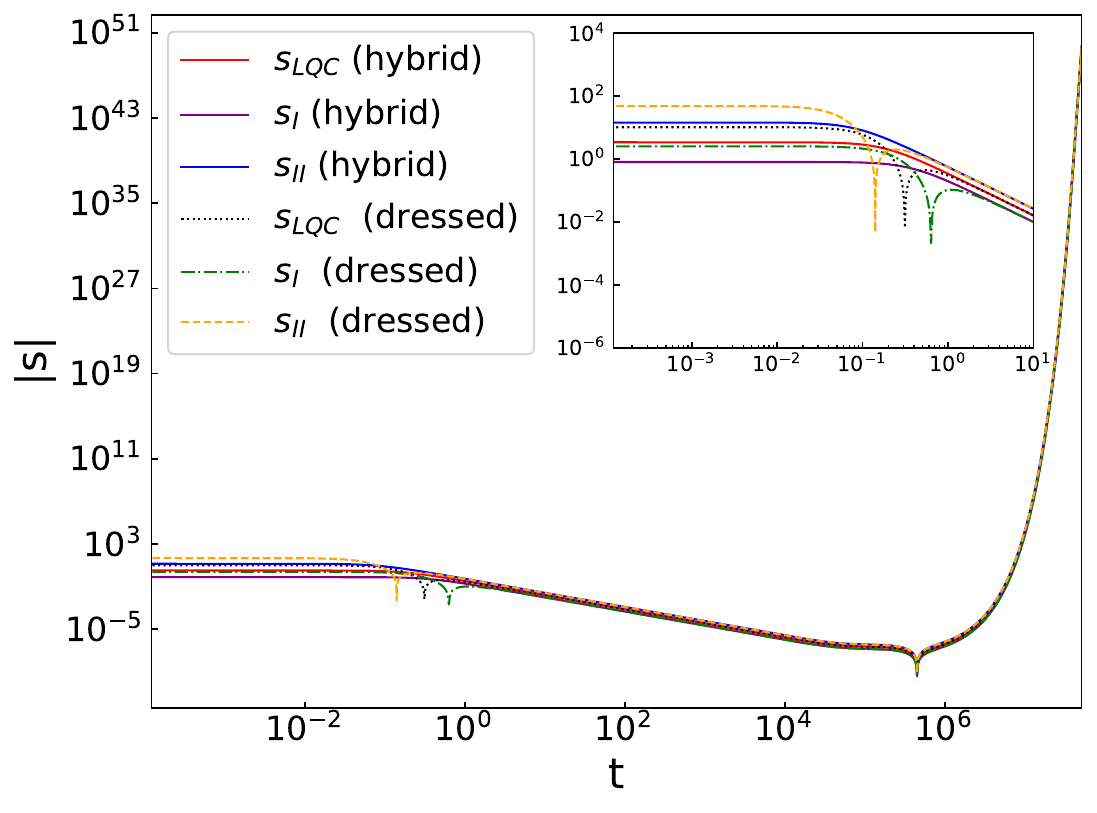} 
    \caption{The left panel compares the absolute value of the effective mass function for LQC ($s_{LQC}$) and mLQC-II ($s_{\scriptscriptstyle{\mathrm{II}}}$) in both the hybrid and the dressed metric approaches in the contracting branch from where the initial states are imposed. The right panel compares the absolute value of the effective mass function for LQC, mLQC-I ($s_{\scriptscriptstyle{\mathrm{I}}}$) and mLQC-II in the expanding branch until $t = 5.1 \times 10^{7}$ (in Planck units).}
    \label{effective-mass-I}
\end{figure}

\subsubsection{Fixation of  the free parameters and the initial conditions}

To obtain the primordial power spectrum, one needs to first fix the background dynamics and then use the appropriate initial states to numerically solve the Mukhanov-Sasaki equation (\ref{MS}). In our simulations, we set the initial conditions for the background dynamics at the bounce point, where the energy density reaches its maximum. Due to the rescaling freedom in volume, we choose $v_{B}=1$ for our numerical solutions without loss of generality. The canonical variable $b$ in each model at the bounce is fixed and takes the value $b_{B} =\pi/2\lambda$ in LQC, $b^{\scriptscriptstyle{\mathrm{I}}}_{B} = \arcsin\left(\sqrt{1/(2 + 2\gamma^2)}\right)/\lambda$ in mLQC-I and $b^{\scriptscriptstyle{\mathrm{II}}}_{B}= \pi/\lambda$ in mLQC-II. Furthermore, the momentum of the scalar field $p_{\phi}$ can be determined by using the effective Hamiltonian constraint while choosing the positive velocity. Therefore, the only free parameters that need to be fixed are the inflaton's mass and the value of the scalar field at the bounce $\phi_{B}$. It turns out that the former controls the amplitude of the primordial power spectrum, and the latter specifies the duration of the inflationary phase, i.e., the number of e-foldings $N_{e}$, or correspondingly, the spectral index of the primordial power spectrum in the almost scale-invariant regime \footnote{The almost scale-invariant regime is where the power spectrum can be approximated by the power law power spectrum, i.e., $P_\mathcal{R} = A_{s} \left({k}/{k_{\star}}\right)^{n_{s}-1}$ where $k_{\star}/a_0 = 0.05~\mathrm{Mpc^{-1}}$, $A_s$ and $n_s$ are given in (\ref{Planck}).}. Given the initial value of the inflaton's mass and $\phi_{B}$ at the bounce, we solve the background equations using the \textsc{solve\_ivp} module from the \textsc{scipy} package in Python, which numerically integrates a system of ordinary differential equations. We use the RK45 method (although the results are the same for different methods), with absolute and relative tolerances set to be $10^{-13}$. 

After fixing the background dynamics, we then proceed with the numerical simulations of the primordial power spectrum. In fact, we set the initial states of the perturbations to be the adiabatic initial states given in (\ref{nu-solution}) in the contracting branch, except for mLQC-I in the dressed metric approach, which is specified by the exact de Sitter solution (\ref{de-sitter-solution}). Given the solutions for the background quantities, we feed these solutions into the Mukhanov-Sasaki equation with the appropriate initial states to compute the primordial power spectrum. To this end, we use the \textsc{pyoscode} package, which has been specifically written to solve second order differential equations with highly oscillatory solutions \cite{Agocs:2019vyk}. This package combines the method of RK45 with the WKB approximation. In fact, when the solution is non-oscillatory, it uses the RK45 method while controlling the error using higher order corrections and skips several cycles using the WKB approximation when the solution is highly oscillatory. This method significantly improves the speed of the code and allows large numbers of simulations. Once the solutions of the mode functions are obtained, we then calculate the primordial power spectrum when the observable modes exit the Hubble horizon (super-horizon) in the slow-roll phase.

\begin{figure}
    \centering
    \includegraphics[scale = 0.4]{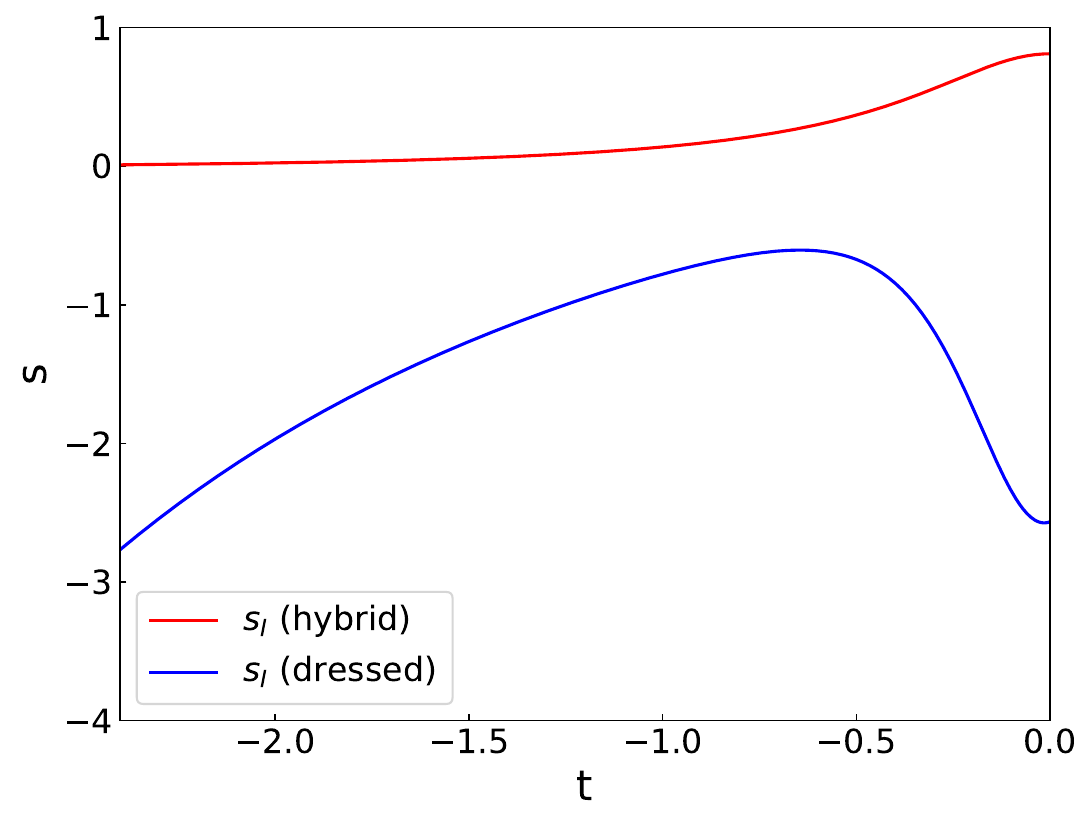}
    \caption{The effective mass function for mLQC-I in the contracting branch from where the adiabatic initial states are imposed for both the hybrid and the dressed metric approaches.}
    \label{effective-mass-II}
\end{figure}

Finally, to compare the angular power spectrum predicted by the loop quantum cosmological models with the angular power spectrum predicted by $\Lambda$CDM best fit to the Planck TT,TE,EE+lowE+lensing data, it is necessary to initially conduct a scale matching for the primordial power spectrum. In fact, we set the scale factor to be unity at the bounce in our numerical simulations, while it is usually fixed to be one at the present time by the Planck mission to find the amplitude and spectral index for the primordial power spectrum in the standard $\Lambda$CDM cosmological model. In order to find the correspondence of the co-moving scales with the observational scales, we fix the amplitude of the curvature power spectrum to be $A_{s}$, i.e., $P_{\mathcal{R}}(k_{\star}) = A_{s}$ to pick some particular co-moving wavenumber of the pivot scale $k_{\star}/a_0 = 0.05~\mathrm{Mpc^{-1}}$. This means that the co-moving wavenumber of the pivot mode obtained in this way depends on the inflaton's mass and the initial value of the scalar field $\phi_{B}$ at the bounce. Moreover, since we solve the exact background equations without using slow-roll approximations, the inflaton's mass obtained from the slow-roll approximation will not result in the central values for the amplitude of the power spectrum and spectral index in Eq. (\ref{Planck}). This means that there is freedom in choosing the pivot mode for which the amplitude of the primordial power spectrum matches the observational data, $A_{s}$, by tuning the inflaton's mass and inflaton field at the bounce in such a way that the inflationary predictions are close to the central values of Eq. (\ref{Planck}).

To restrict above freedom and place each model's predictions on an equal footing for later comparison, we choose the inflaton's mass and inflaton field at the bounce while the amplitude of the primordial power spectrum matches the $A_{s}$ for a particular $k_{\star}$, which is well inside the scale invariant regime. Hence, this will guarantee that the relevant angular power spectrum completely matches the angular power spectrum predicted by $\Lambda$CDM best fit to the Planck TT,TE,EE+lowE+lensing data at large $l$ multipoles and has the smallest deviation at large angles, i.e., low $l$ multipoles for each model. Additionally, as we will see, different regularizations and quantum ambiguities modify the infrared and the intermediate regimes in the power spectrum while producing the same scale-invariant regime. Hence, in order to compare these models appropriately, we finely tune the inflaton's mass and $\phi_{B}$ in such a way that all three models, LQC and mLQCs, not only predict the same number of e-folding, i.e., $N_{e} \simeq 65$, but also lead to the relative difference in the amplitude of the power spectrum being less than one percent in the scale invariant regime. It should be mentioned that while the analysis presented in this manuscript focused on these number of e-foldings, our results did not change when the e-foldings were changed to $60$ or $70$. It turns out that in this case, there is not much freedom in choosing a quite different pivot mode for different models, which might affect our conclusion. To achieve this, we choose the inflaton's mass to be $m = 2.7 \times 10^{-6}$, which is different from what is obtained from slow-roll approximations, i.e., $m= 2.44\times 10^{-6}$, and the value of the inflaton field at the bounce is fixed for each model accordingly to achieve $N_{e} = 65$. We again note that the inflaton's mass is slightly different from what is obtained from slow-roll approximation because we numerically solve exact equations, and that is not because of quantum gravity effects. In this sense, we can compare the primordial power spectrum with different regularizations and quantum ambiguities in the infrared and intermediate regimes which can result in modifications in the angular power spectrum. In the following we aim to understand how the amplitude and shape of the primordial power spectrum in the infrared and the intermediate regimes would change for different regularizations, quantum ambiguities, initial states, and where the initial states are imposed. 

\begin{figure}
    \centering
    \includegraphics[scale = 0.34]{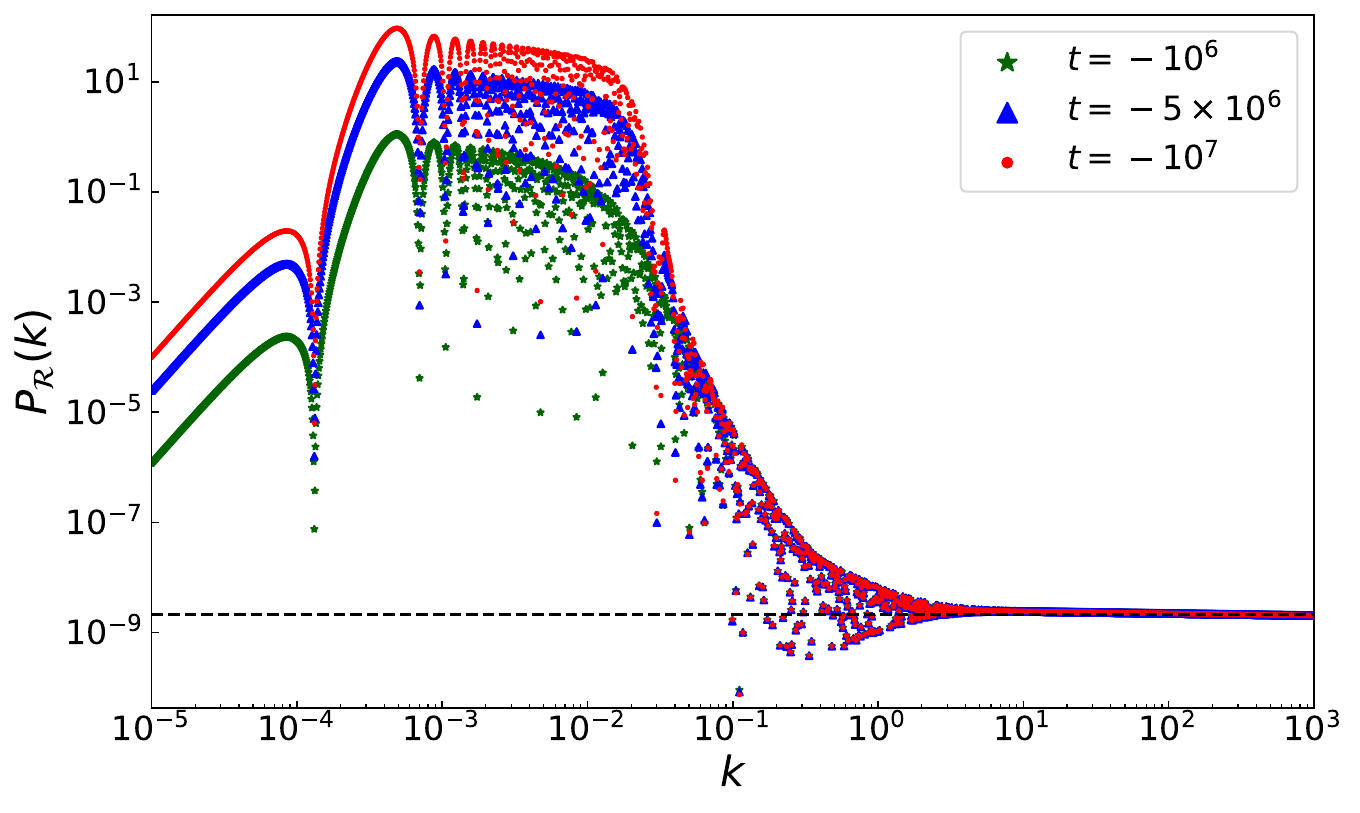} \ \ \ \ \ \includegraphics[scale = 0.34]{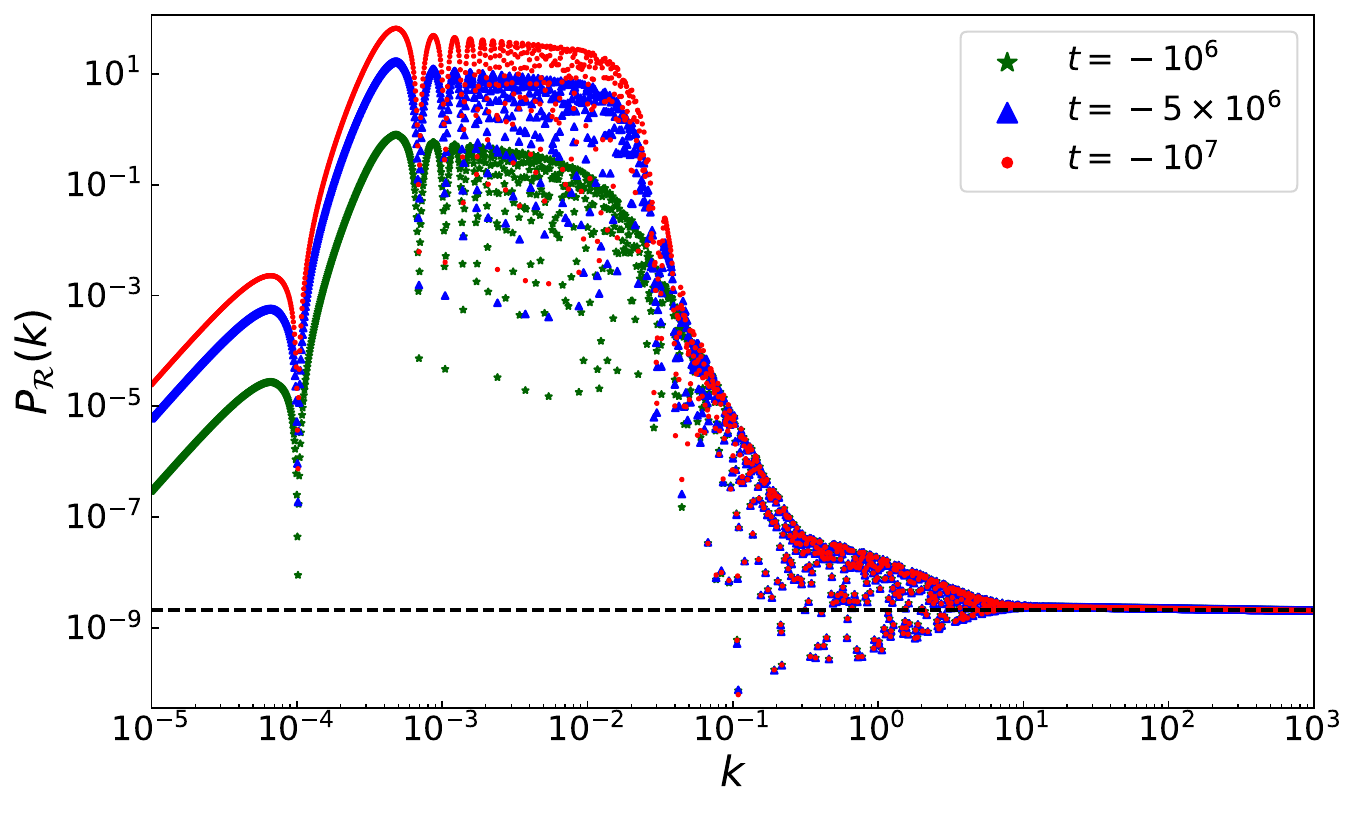}
    \caption{The primordial power spectrum for LQC with $\phi_{B} = -1.4306$ and $m = 2.7 \times 10^{-6}$, while second order adiabatic initial states are imposed at $t = -10^{6}$, $-5\times 10^{6}$, and $-10^7$ for the hybrid (left) and the dressed metric (right) approaches (all in Planck units). The corresponding $k_{\star}$ for imposing adiabatic initial states at $t = -10^{6}$, $-5\times 10^{6}$, and $-10^7$ in the hybrid approach (left) are ${k_{\star}} = 492.754$, ${k_{\star}} = 491.134$, and ${k_{\star}} = 490.459$, and for the dressed metric approach (right) are ${k_{\star}} = 92.365$, ${k_{\star}} = 491.134$, and ${k_{\star}} = 490.334$, respectively. The dashed line is the central value for the amplitude of the primordial power spectrum, i.e., $A_{s} = 2.0989 \times 10^{-9}$, according to the Planck collaboration in the TT, TE, EE-lowE+lensing 68\% limits data at pivot scale $k_\star/a_0 = 0.05 \mathrm{Mpc^{-1}}$.}
    \label{different-initial-time}
\end{figure}

\begin{figure}
    \centering
    \includegraphics[scale = 0.34]{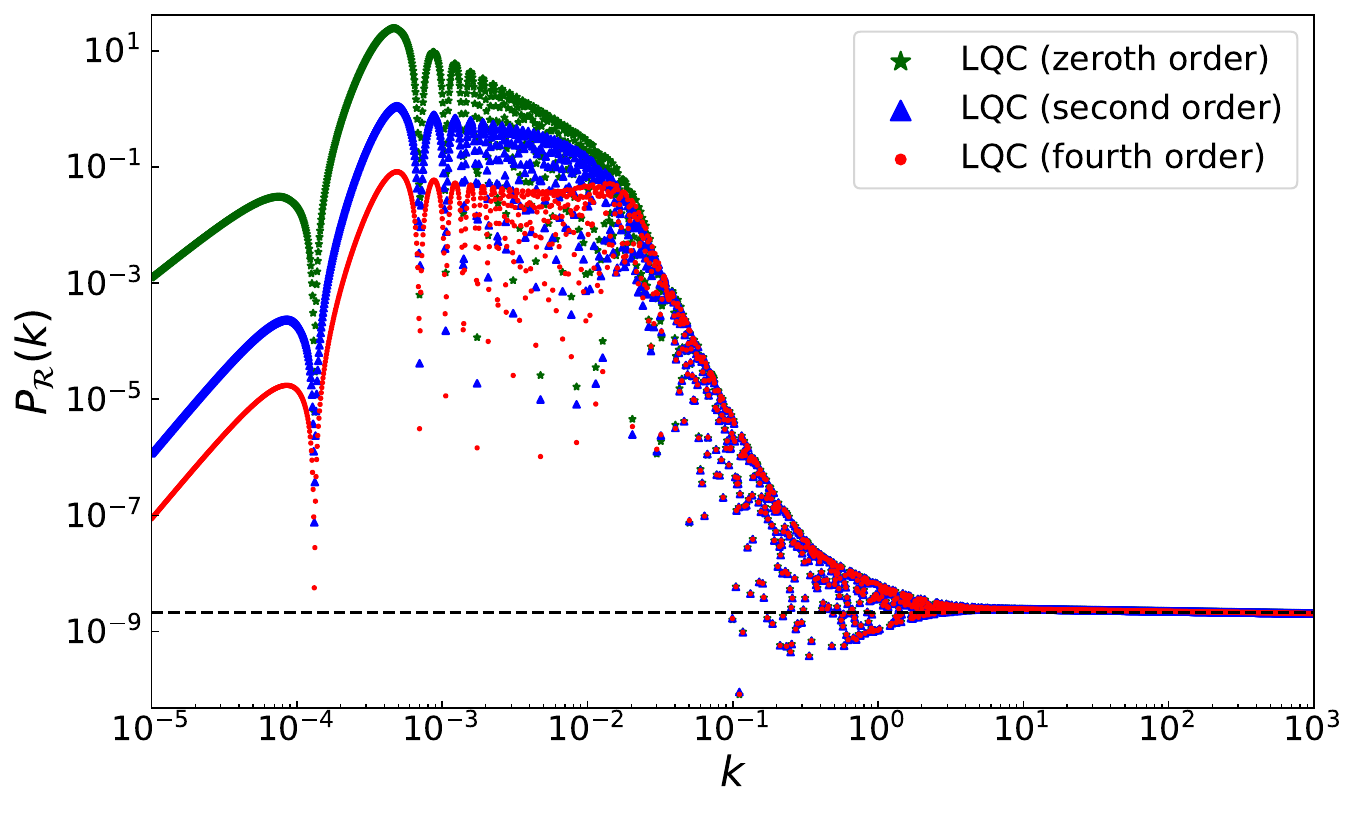} \ \ \ \ \ \includegraphics[scale = 0.34]{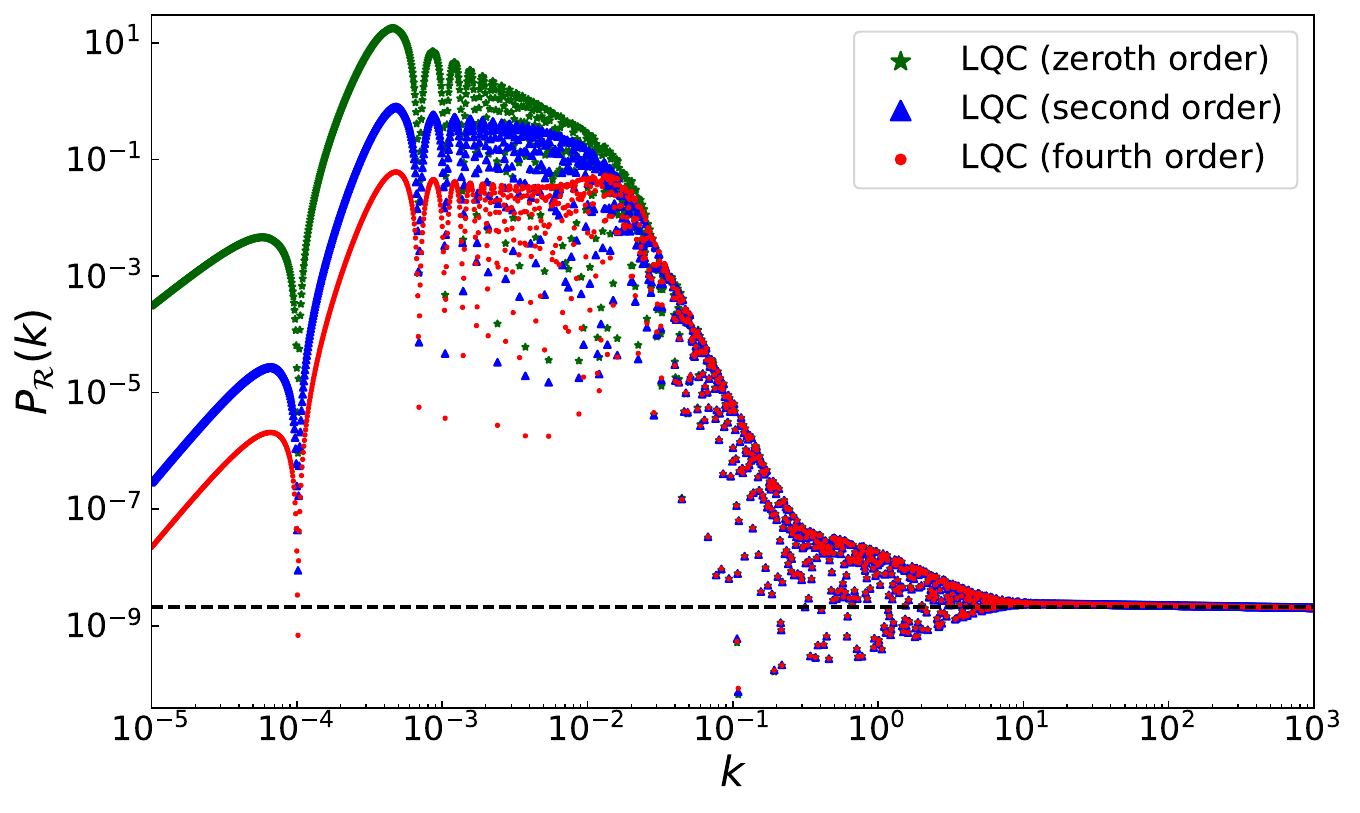}
    \caption{The primordial power spectrum for LQC with different adiabatic initial states (zeroth, second, and fourth order) in the hybrid approach (left panel) and the dressed metric (right panel) approach, while $t = -10^{6}$, $\phi_{B} = -1.4306$, and $m = 2.7 \times 10^{-6}$ (all in Planck units). The corresponding $k_{\star}$ for zeroth, second, and fourth order adiabatic initial states in the hybrid approach (left panel) are ${k_{\star}^{(0)}} = 493.396$, ${k_{\star}^{(2)}} = 492.754$, and ${k_{\star}^{(4)}}= 490.168$ where the upper indices $0$, $2$, and $4$ denote the order of adiabatic initial states, and for the dressed metric approach (right panel) are ${k_{\star}^{(0)}} = 490.297$, ${k_{\star}^{(2)}} = 492.365$, and ${k_{\star}^{(4)}}= 494.283$, respectively. The dashed line is the central value for the amplitude of the primordial power spectrum, i.e., $A_{s} = 2.0989 \times 10^{-9}$, according to the Planck collaboration in the TT, TE, EE-lowE+lensing 68\% limits data.}
    \label{power-spectrum-LQC}
\end{figure}

\subsubsection{Comparison of the effective mass functions in different models and approaches}

As we have discussed earlier, the effects of different regularizations and quantum ambiguities are encoded in the time-dependent effective mass function. To this end, the effective mass functions of LQC and mLQC-II in the contracting branch from where the adiabatic initial states are imposed are shown in the left panel of Fig. \ref{effective-mass-I} and also in the expanding branches (including effective mass function for mLQC-I) in the right panel of Fig. \ref{effective-mass-I} for both the hybrid and the dressed metric approaches. From these figures, it is obvious that the effective mass function far away from the bounce in the hybrid and the dressed metric approaches are the same for both LQC and mLQC-II. Moreover, the order of magnitude of the effective mass function is also the same for both LQC and mLQC-II. However, from the right panel in Fig. \ref{effective-mass-I}, one can see the noticeable differences among different models near the bounce. From the left panel in Fig. \ref{effective-mass-I}, one can see that the effective mass function is positive where the adiabatic initial states are imposed for both LQC and mLQC-II. In Fig. \ref{effective-mass-II}, we plot the effective mass function for mLQC-I in both the hybrid (red curve) and the dressed metric (blue curve) approaches. From this plot, one can see that the effective mass function in the dressed metric approach is negative in the contracting branch where the exact de Sitter solution (\ref{de-sitter-solution}) is imposed, and it also goes to a large negative value by further going into the contracting branch. On the other hand, the effective mass function for the hybrid approach goes to very large positive values by further going into the contracting branch.

\subsubsection{The influence of different initial times on the primordial power spectrum}

In this subsection, we study the effects of choosing different initial times for setting the initial states in the contracting phase on the shape of the primordial power spectrum. As an illustrative example, we plot the primordial power spectrum for LQC with $\phi_{B} = -1.4306$ and $m = 2.7\times 10^{-6}$ in the range of the co-moving wave number $k \in (10^{-5}, 1000)$ while the second order adiabatic initial states are imposed at $t = -10^{6}$, $-5 \times 10^{6}$ and $-10^{7}$ for both the hybrid (left) and the dressed metric (right) approaches in Fig. \ref{different-initial-time}. In this plot, the green star is the primordial power spectrum for adiabatic initial states imposed at $t = -10^6$, the blue triangle is for $t = -5 \times 10^6$, and the red dotted is for $t = -10^7$. The corresponding $k_{\star}$ for imposing adiabatic initial states at $t = -10^{6}$, $-5\times 10^{6}$, and $-10^7$ in the hybrid approach (left) are ${k_{\star}} = 493.396$, ${k_{\star}} = 492.134$, and ${k_{\star}} = 491.459$, and for the dressed metric approach (right) are ${k_{\star}} = 490.297$, ${k_{\star}} = 491.134$, and ${k_{\star}} = 492.334$, respectively. We should also point out that there are $2000$ samples for each primordial power spectrum in the figure. From these figures, one can see that the primordial power spectrum in LQC and its modified version, as we will see later, can be generally divided into three distinctive regimes: the suppressed infrared regime for $k = (10^{-5}, 10^{-4})$, the amplified oscillatory regime for $k = (10^{-4}, 1)$, and the scale-invariant regime for $k = (1, 1000)$. From these plots, one can clearly see that the amplification of the power spectrum in the intermediate regime depends on how far from the bounce the adiabatic initial states are imposed in the contracting branch. In fact, as the adiabatic initial states are imposed further in the contracting branch, the primordial power spectrum has a larger amplification in the intermediate regime. This behavior is also observed in the case of mLQC-I for the hybrid approach and mLQC-II for both the hybrid and the dressed metric approaches with zeroth, second, and fourth order adiabatic initial states, except for the case of mLQC-I in the dressed metric approach, where the initial state is specified using the exact de Sitter solution. This is mainly because adiabatic initial states are just the approximate solutions of the Mukhanov-Sasaki equation of the mode function. In contrast, in the dressed metric approach of mLQC-I, the exact solution of the Mukhanov-Sasaki equation, which is the de Sitter initial state, is available, and the resulting primordial power spectrum is then independent of the initial time. In the case of mLQC-I, as the adiabatic initial states are imposed further in the contracting branch, the scale-invariant regime occurs in larger wave modes in comparison with LQC and mLQC-II due to the special properties of the effective mass function in this model, which is why the chosen initial time for mLQC-I is very different from the initial time for LQC and mLQC-II. In fact, to have the same inflationary predictions compared with two other models, we must choose the initial time to have different values from the other two models. Otherwise, the relative difference in the primordial power spectrum in the scale-invariant regime will not be less than one percent when it is compared with two other models. 

\begin{figure}
    \centering
    \includegraphics[scale = 0.34]{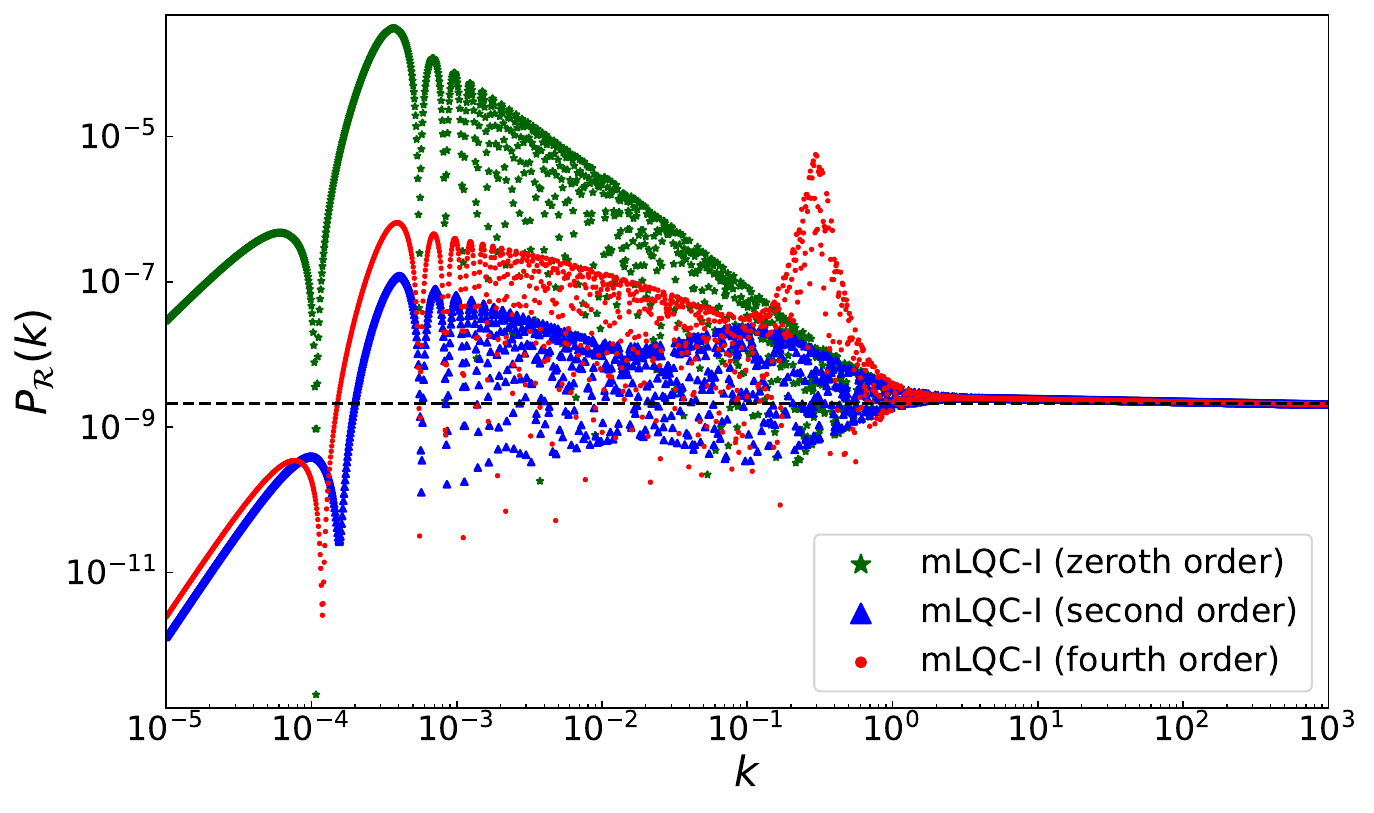}  \ \ \ \ \ \includegraphics[scale = 0.34]{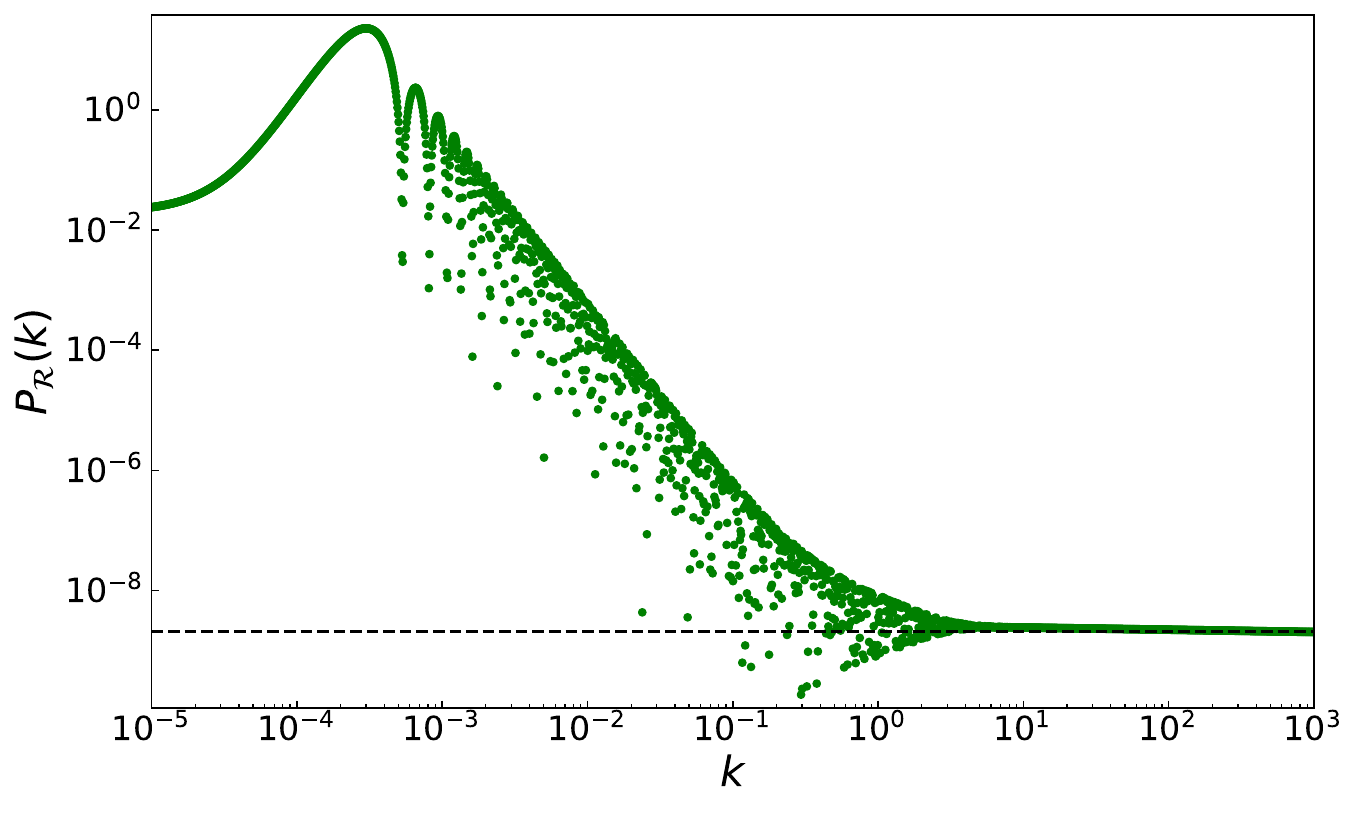} 
    \caption{The primordial power spectrum for mLQC-I with different adiabatic initial states (zeroth, second, and fourth order) in the hybrid approach (left) and with the de Sitter initial state in the dressed metric approach (right) while $t = -2.4$, $\phi_{B} = -1.31$ and $m = 2.7 \times 10^{-6}$ (in Planck units). The corresponding $k_{\star}$ for zeroth, second, and fourth order adiabatic initial states in the hybrid approach (left) are ${k_{\star}^{(0)}}= 518.996$, ${k_{\star}^{(2)}}= 518.692$, and ${k_{\star}^{(4)}}= 519.114$ where the upper indices $0$, $2$, and $4$ denote the order of adiabatic initial states, and for the dressed metric approach is ${k_{\star}^{\mathrm{(ds)}}} = 520.332$ (ds denotes de Sitter  initial state), respectively. The dashed line is the central value for the amplitude of the primordial power spectrum, i.e., $A_{s} = 2.0989 \times 10^{-9}$, according to the Planck collaboration in the TT, TE, EE-lowE+lensing 68\% limits data.}
    \label{power-spectum-mLQC-I}
\end{figure}

\begin{figure}
    \centering
    \includegraphics[scale = 0.34]{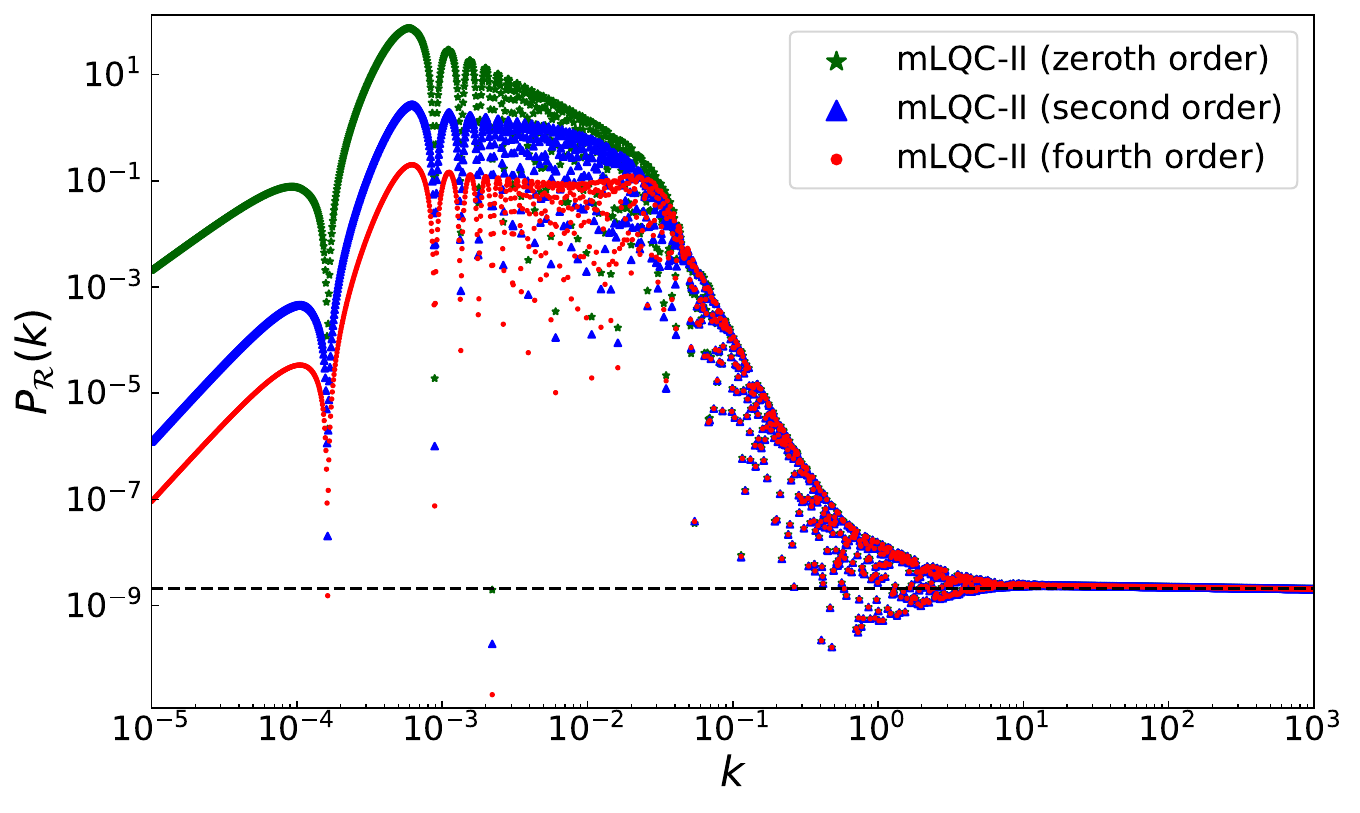}\ \ \ \ \ \includegraphics[scale = 0.34]{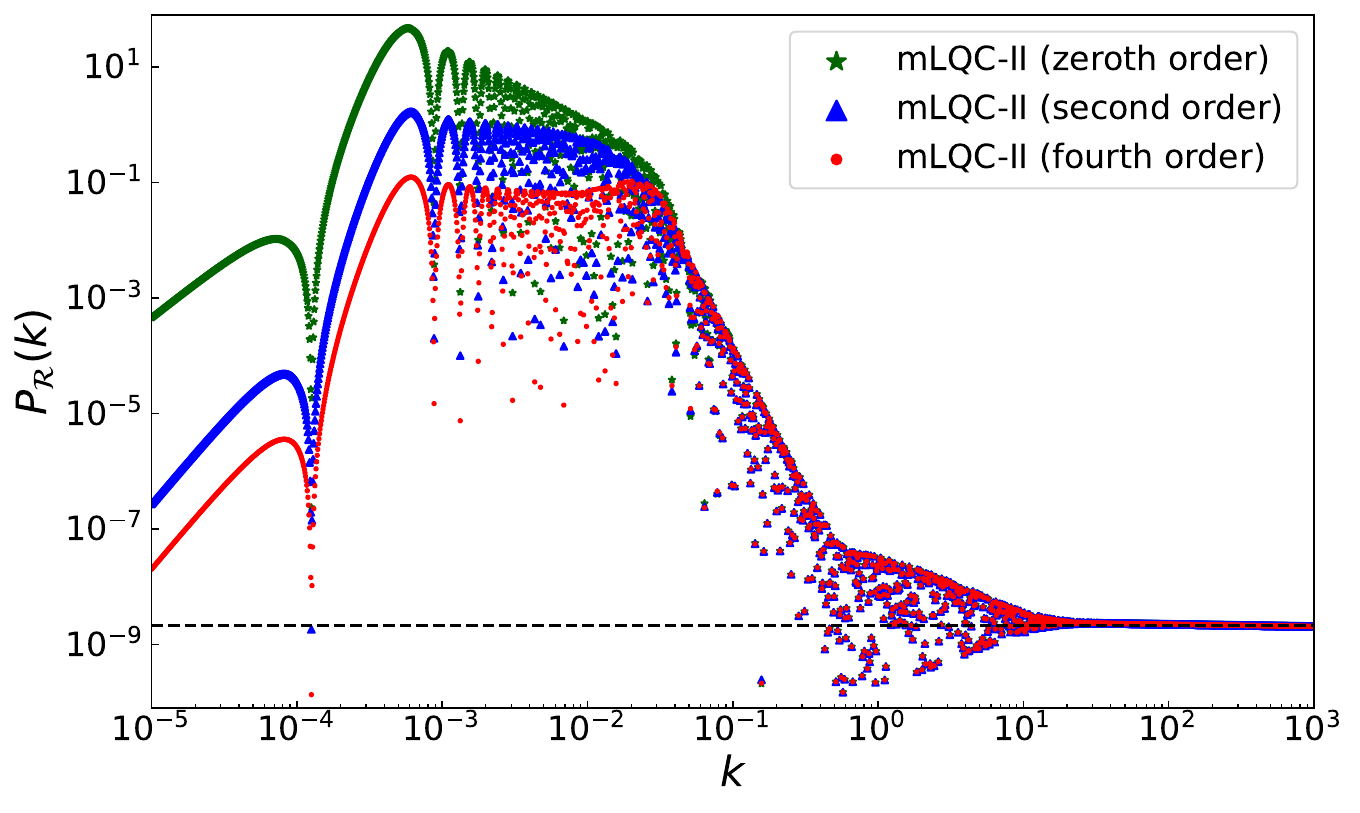}
    \caption{The primordial power spectrum for mLQC-II with different adiabatic initial states (zeroth, second, and fourth order) in the hybrid approach (left) and the dressed metric (right) approaches while $t = -10^{6}$, $\phi_{B} = -1.54$, and $m = 2.7 \times 10^{-6}$. The corresponding $k_{\star}$ for zeroth, second, and fourth order adiabatic initial states in the hybrid approach (left) are ${k_{\star}^{(0)}} = 478.882$, ${k_{\star}^{(2)}} = 478.500$, and ${k_{\star}^{(4)}} = 480.513$ where the upper indices $0$, $2$, and $4$ denote the order of adiabatic initial states, and for the dressed metric approach are ${k_{\star}^{(0)}}= 479.876$, ${k_{\star}^{(2)}}= 479.228$, and ${k_{\star}^{(4)}}= 480.008$, respectively. The dashed line is the central value for the amplitude of the primordial power spectrum, i.e., $A_{s} = 2.0989 \times 10^{-9}$, according to the Planck collaboration in the TT, TE, EE-lowE+lensing 68\% limits data.}
    \label{power-spectrum-mLQC-II}
\end{figure}

\subsubsection{The influence of different initial states on the primordial power spectrum }

With the given initial conditions for the background dynamics and the initial states for the linear perturbations, we numerically compute the primordial power spectrum in both the dressed metric approach and the hybrid approach for all three models and present these results in Fig. \ref{power-spectrum-LQC}-\ref{power-spectrum-mLQC-II}. In Fig. \ref{power-spectrum-LQC}, we plot the primordial power spectrum for LQC with $\phi_{0} = -1.4306$ and $m = 2.7\times 10^{-6}$ while adiabatic initial states (zeroth, second, and fourth order) are imposed at $t = -10^{6}$ for both the hybrid (left) and the dressed metric (right) approaches. The dashed line denotes the central value for the amplitude of the primordial power spectrum, i.e., $A_{s} = 2.0989 \times 10^{-9}$, according to the Planck collaboration in the TT,TE,EE-lowE+lensing 68\% limits data at pivot scale $k_\star/a_0 = 0.05 \mathrm{Mpc^{-1}}$. The corresponding $k_{\star}$ for zeroth, second, and fourth order adiabatic initial states in the hybrid approach (left) are ${k_{\star}^{(0)}}=493.396$, ${k_{\star}^{(2)}}=492.754$ and ${k_{\star}^{(4)}}=490.168$ where the upper indices $0$, $2$, and $4$ denote the order of adiabatic initial states, and for the dressed metric approach (right) are ${k_{\star}^{(0)}}= 490.297$, ${k_{\star}^{(2)}} = 492.365$, and ${k_{\star}^{(4)}} = 494.283$, respectively. As it is obvious from both panels, the amplification of the primordial power spectrum in the intermediate regime depends on the order of adiabatic initial states in both the hybrid and the dressed metric approaches. In fact, for higher order adiabatic initial states, the power spectrum has smaller amplification in the intermediate regime in both the hybrid and the dress metric approaches. Moreover, in a very small $k$ regime (the infrared regime), the suppression of the primordial power spectrum is slightly stronger in the case of the dressed metric approach. Apart from that, from the primordial power spectrum, it is not easy to distinguish the dressed metric approach from the hybrid approach, where the amplitude of the primordial power spectrum in the intermediate regime in two approaches looks close to each other once the same initial states are chosen. 

In Fig. \ref{power-spectum-mLQC-I}, we plot the primordial power spectrum for mLQC-I with $\phi_{0}=-1.31$ and $m=2.7\times 10^{-6}$ while initial states are imposed at $t=-2.4$ for both the hybrid (left) and the dressed metric (right) approaches. In this case, we use zeroth, second, and fourth order adiabatic initial states for the hybrid approach but use the de Sitter initial state for the dressed metric approach, which is why only one primordial power spectrum is plotted in the right panel of Fig. \ref{power-spectum-mLQC-I}. The corresponding $k_{\star}$ for zeroth, second, and fourth order adiabatic initial states in the hybrid approach (left) are ${k_{\star}^{(0)}}=518.996$, ${k_{\star}^{(2)}}=518.692$, and ${k_{\star}^{(4)}} = 519.114$, and for the dressed metric approach is ${k_{\star}^{(ds)}}=520.332$ (ds denotes de Sitter initial state), respectively. From these figures, it is obvious that the amplitude of the intermediate regime of the primordial power spectrum is larger for the dressed metric approach in comparison with the hybrid approach. Moreover, in the case of the dressed metric approach, the primordial power spectrum monotonically increases in the intermediate regime and then becomes constant at a very small $k$ regime (the infrared regime). In the case of the hybrid approach, one can see that the amplification of the primordial power spectrum in the intermediate regime depends on the order of adiabatic initial states. However, there are important differences between LQC and mLQC-I in this case. First, the amplification in the intermediate regime for mLQC-I is smaller in comparison to LQC for all zeroth, second, and fourth order adiabatic initial states in the hybrid approach. Second, for fourth order adiabatic initial states, there seems to be a peak of large magnitude in the rightmost part of the intermediate regime, which precedes the scale-invariant regime. 

Finally, we plot the primordial power spectrum for mLQC-II with $\phi_{0} = -1.54$ and $m = 2.7\times 10^{-6}$ while adiabatic initial states (zeroth, second, and fourth order) are imposed at $t= -10^{6}$ for both the hybrid (left) and the dressed metric (right) approaches in Fig. \ref{power-spectrum-mLQC-II}. The corresponding $k_{\star}$ for zeroth, second, and fourth order adiabatic initial states in the hybrid approach (left) are ${k_{\star}^{(0)}}=478.882$, ${k_{\star}^{(2)}}=478.500$, and ${k_{\star}^{(4)}}=480.513$, and for the dressed metric approach are ${k_{\star}^{(0)}}=479.876$, ${k_{\star}^{(2)}}=479.228$, and ${k_{\star}^{(4)}}=480.008$, respectively. The behavior of the primordial power spectrum is very similar to LQC in both the hybrid and the dressed metric approaches. However, the primordial power spectrum has a larger amplification in the intermediate regime for mLQC-II in comparison to LQC. As we will see, these differences in the intermediate regime lead to different modifications in the angular power spectrum at large angles, i.e., low $l$ multipoles.

\subsection{The angular power spectra in loop quantum cosmological models} \label{sectionIV-B}

In order to compute the angular power spectrum, we feed the primordial power spectrum computed in Figs. \ref{power-spectrum-LQC}-\ref{power-spectrum-mLQC-II} into CAMB code as an external primordial power spectrum. In fact, since the inflationary phase is occurring far away from the bounce regime, the background dynamics are the same as in the classical cosmology when inflation begins, while the effect of the pre-inflationary phase is encoded in the initial states and also the effective mass function in the Mukhanov-Sasaki equation. Hence, we can use the transfer function computed by CAMB for the standard cosmology in this case, while using the primordial power spectrum computed in the previous section. Before feeding the primordial power spectrum into the CAMB code, we first take an average over 20 samples to make the power spectrum smoother, then we use scale matching (explained earlier) to normalize the primordial power spectrum at $k_{\star}/a_0 = 0.05\mathrm{Mpc^{-1}}$ to be able to compare the prediction of the model with observational data. Therefore, given the primordial power spectrum for all three models with different initial states, we calculate the angular power spectrum in both the hybrid and the dressed metric approaches.

\begin{figure}
    \centering
    \includegraphics[width=12cm]{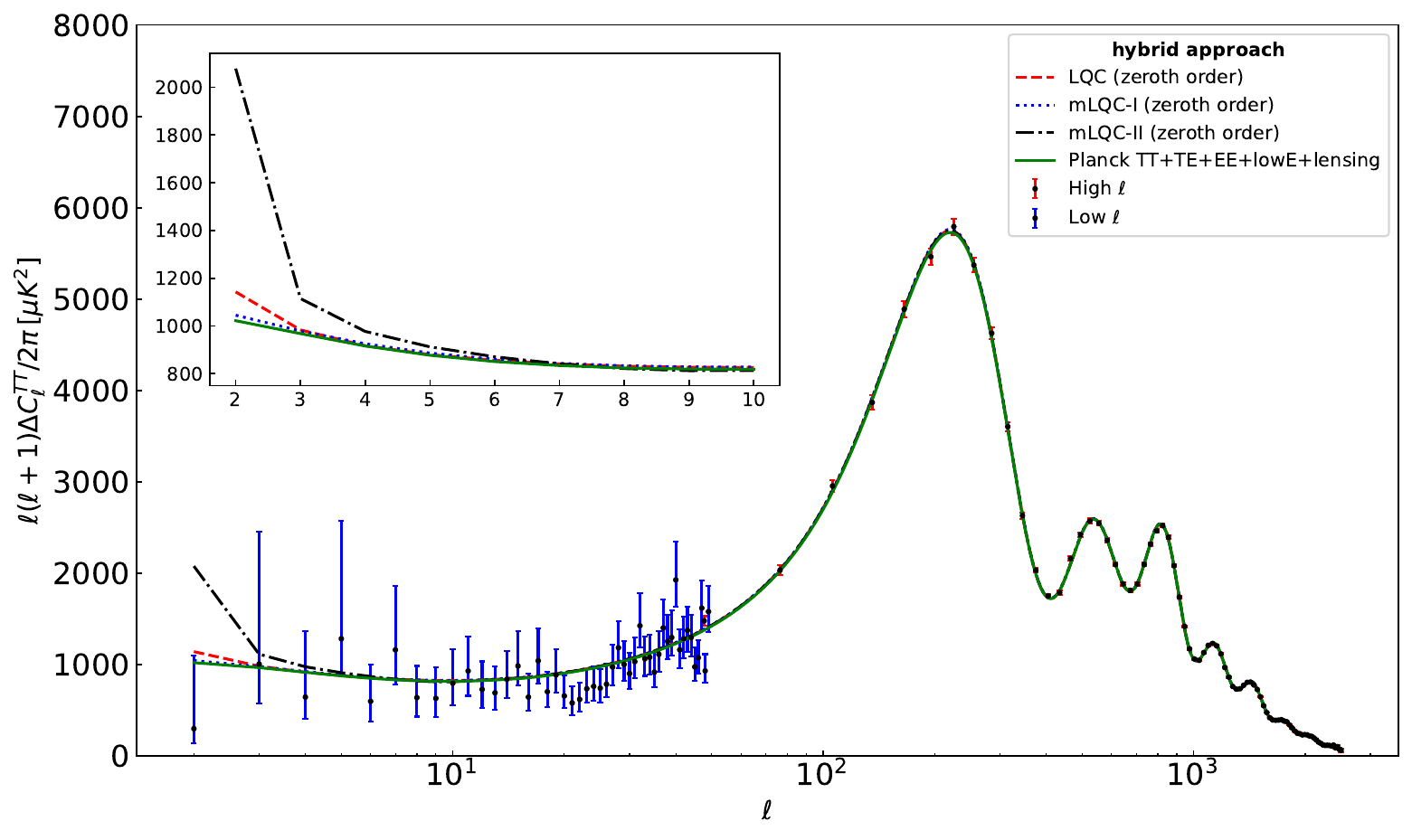}  
    \caption{The angular power spectrum predicted by LQC models in the case of the hybrid approach for zeroth order adiabatic initial state. The black dots are the Planck 2018 temperature angular power spectrum, with blue error bars for low $l$ multipoles and red error bars for large $l$ multipoles. The green curve is the $\Lambda$CDM angular power spectrum best fit to Planck 2018 data.}
    \label{angular-power-zero-order-hybrid}
\end{figure}

\begin{figure}
    \centering
 \includegraphics[width=12cm]{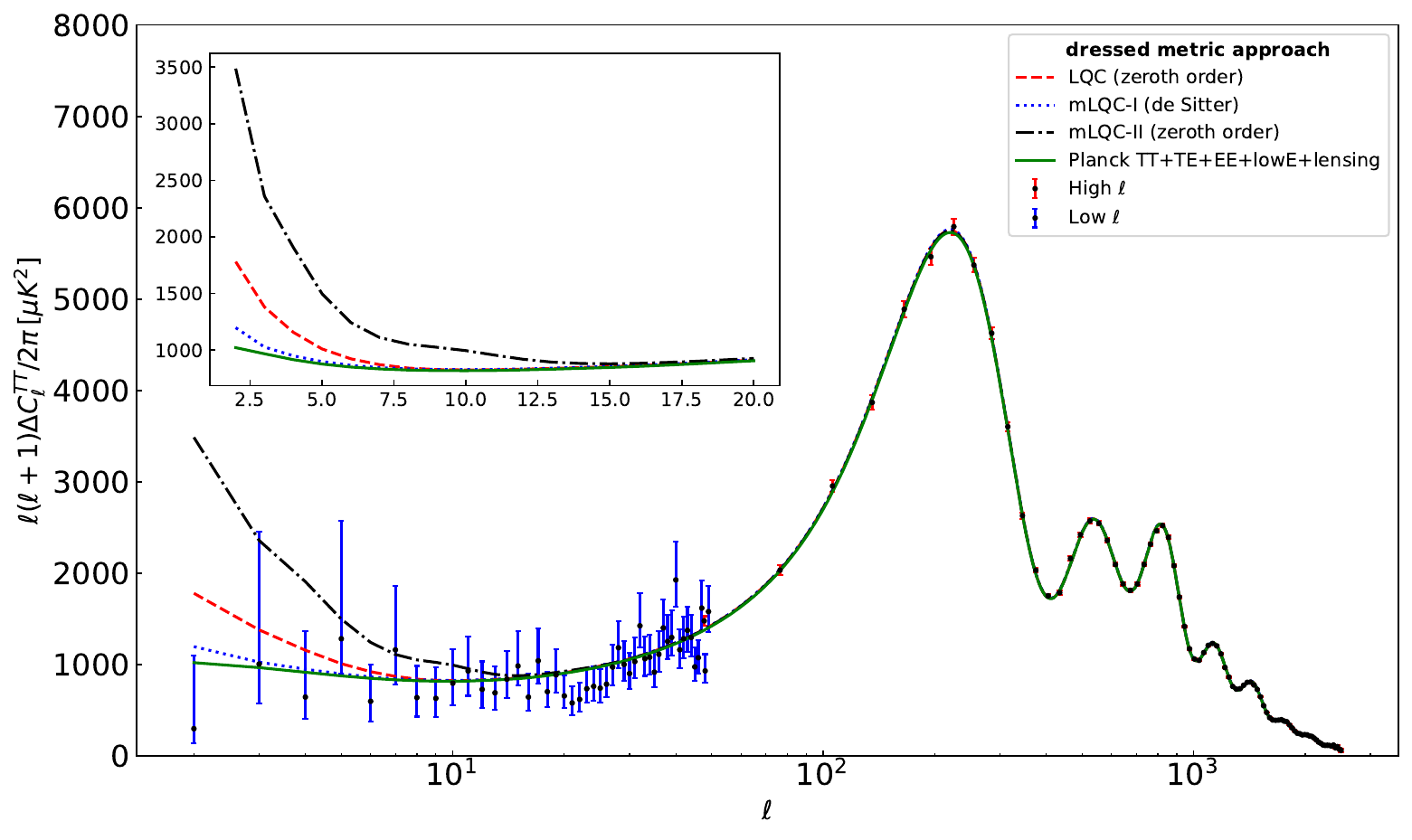}
    \caption{The angular power spectrum predicted by LQC models in the case of the dressed metric approach, while zeroth order adiabatic initial states are used for LQC and mLQC-II and de Sitter initial state is used for mLQC-I. The black dots are the Planck 2018 temperature angular power spectrum, with blue error bars for low $l$ multipoles and red error bars for large $l$ multipoles. The green curve is the $\Lambda$CDM angular power spectrum best fit to Planck 2018 data.}
    \label{angular-power-zero-order-dressed}
\end{figure}

\begin{figure}
    \centering
    \includegraphics[width=12cm]{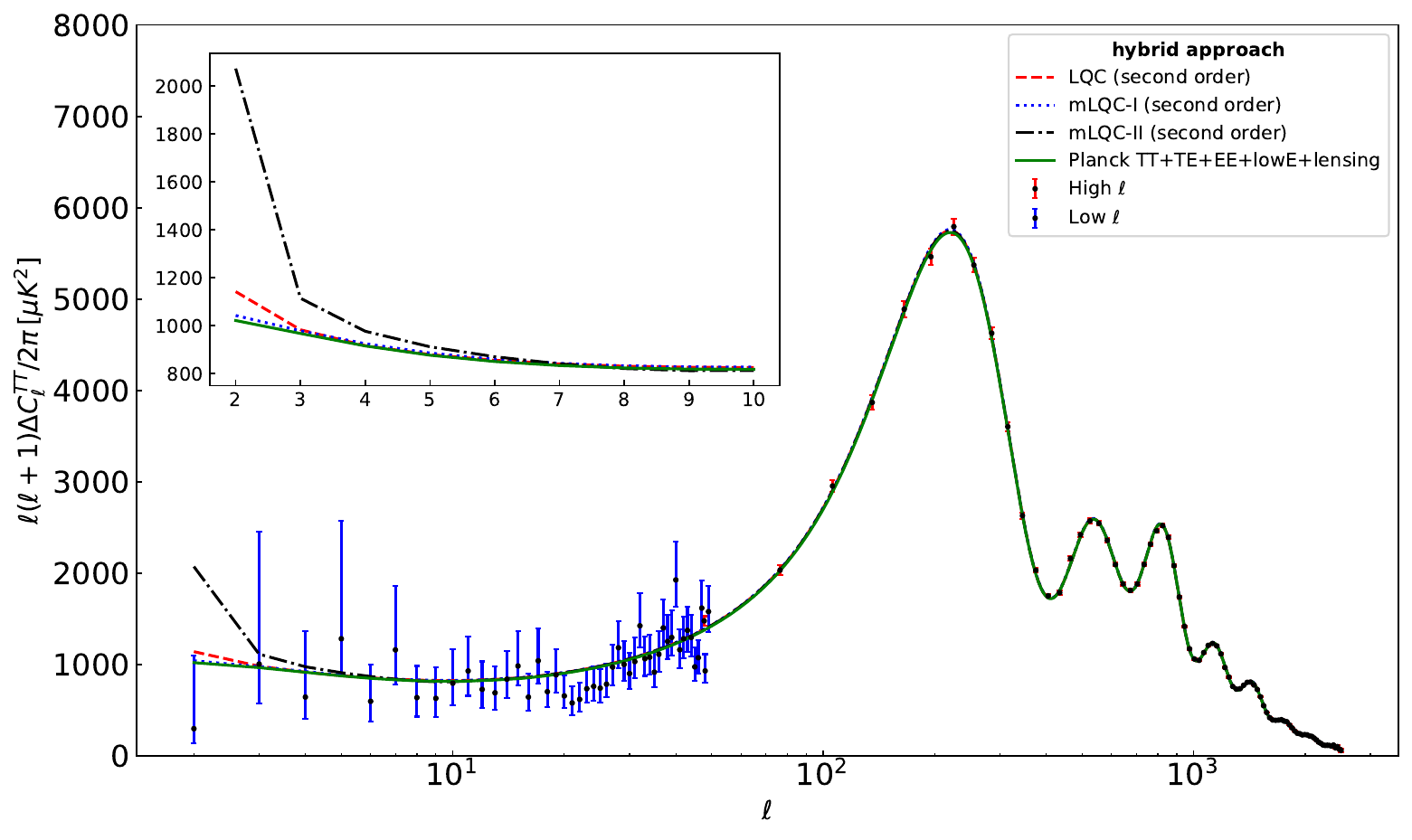}
     \caption{The angular power spectrum predicted by LQC models in the case of the hybrid approach for second order adiabatic initial states. The black dots are the Planck 2018 temperature angular power spectrum, with blue error bars for low $l$ multipoles and red error bars for large $l$ multipoles. The green curve is the $\Lambda$CDM angular power spectrum best fit to Planck 2018 data.}
    \label{angular-power-second-order-hybrid}
\end{figure}

\begin{figure}
    \centering
   \includegraphics[width=12cm]{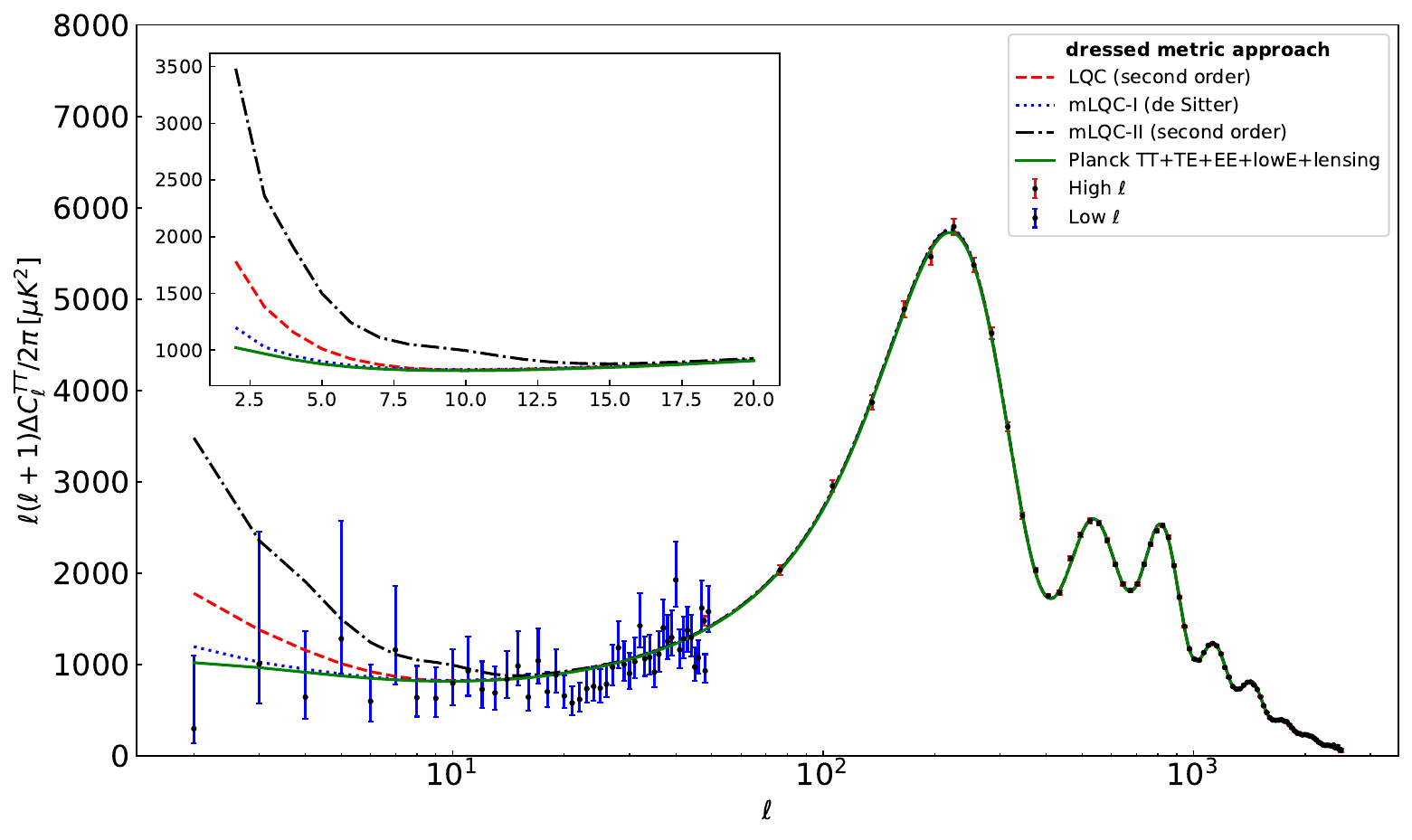}
     \caption{The angular power spectrum predicted by LQC models in the case of the dressed metric approach, while second order adiabatic initial states are used for LQC and mLQC-II and de Sitter initial state is used for mLQC-I. The black dots are the Planck 2018 temperature angular power spectrum, with blue error bars for low $l$ multipoles and red error bars for large $l$ multipoles. The green curve is the $\Lambda$CDM angular power spectrum best fit to Planck 2018 data.}
    \label{angular-power-second-order-dressed}
\end{figure}

\begin{figure}
    \centering
    \includegraphics[width=12cm]{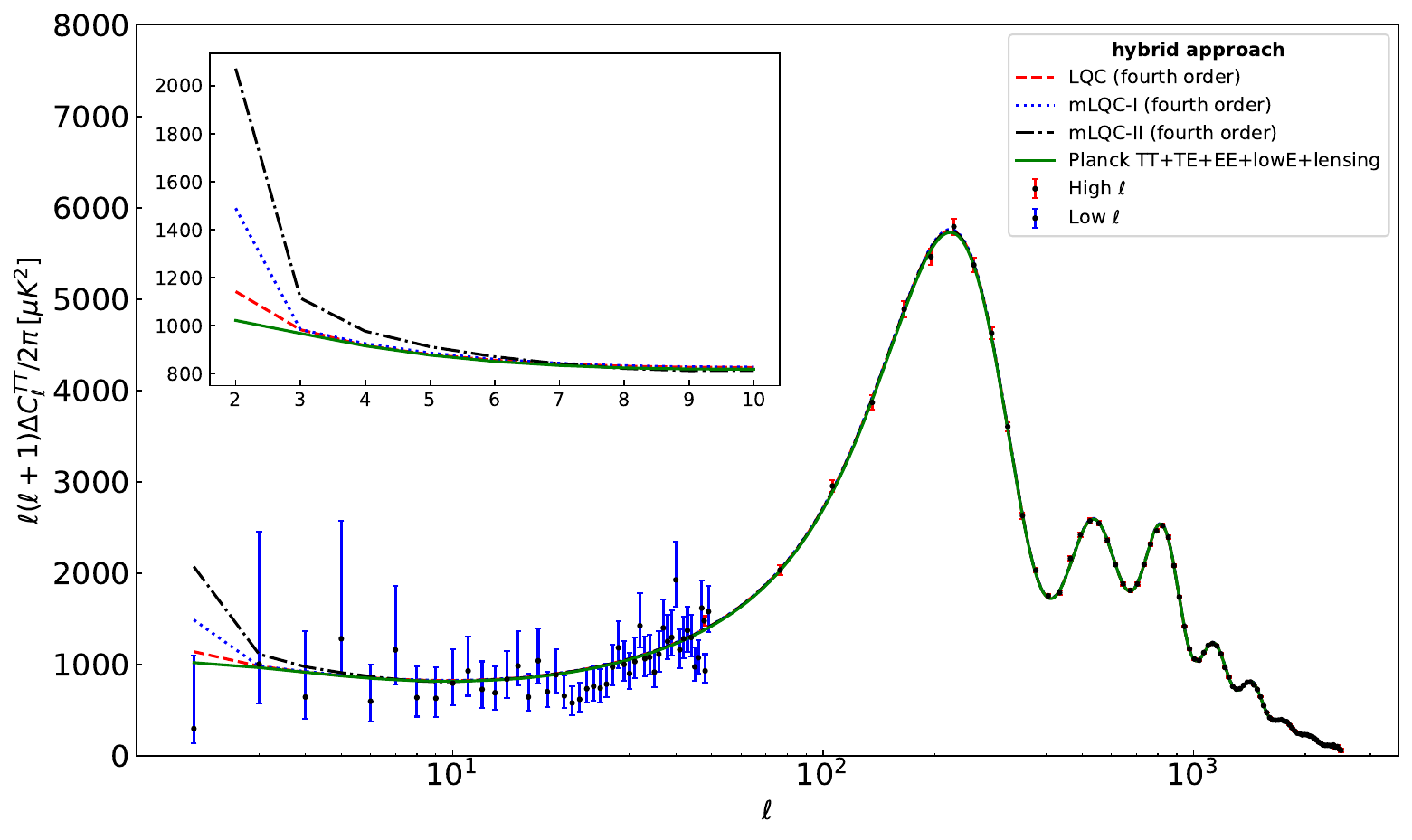} 
     \caption{The angular power spectrum predicted by LQC models in the case of the hybrid approach for the fourth order adiabatic initial states. The black dots are the Planck 2018 temperature angular power spectrum, with blue error bars for low $l$ multipoles and red error bars for large $l$ multipoles. The green curve is the $\Lambda$CDM angular power spectrum best fit to Planck 2018 data.}
    \label{angular-power-fourth-order-hybrid}
\end{figure}

\begin{figure}
    \centering
    \includegraphics[width=12cm]{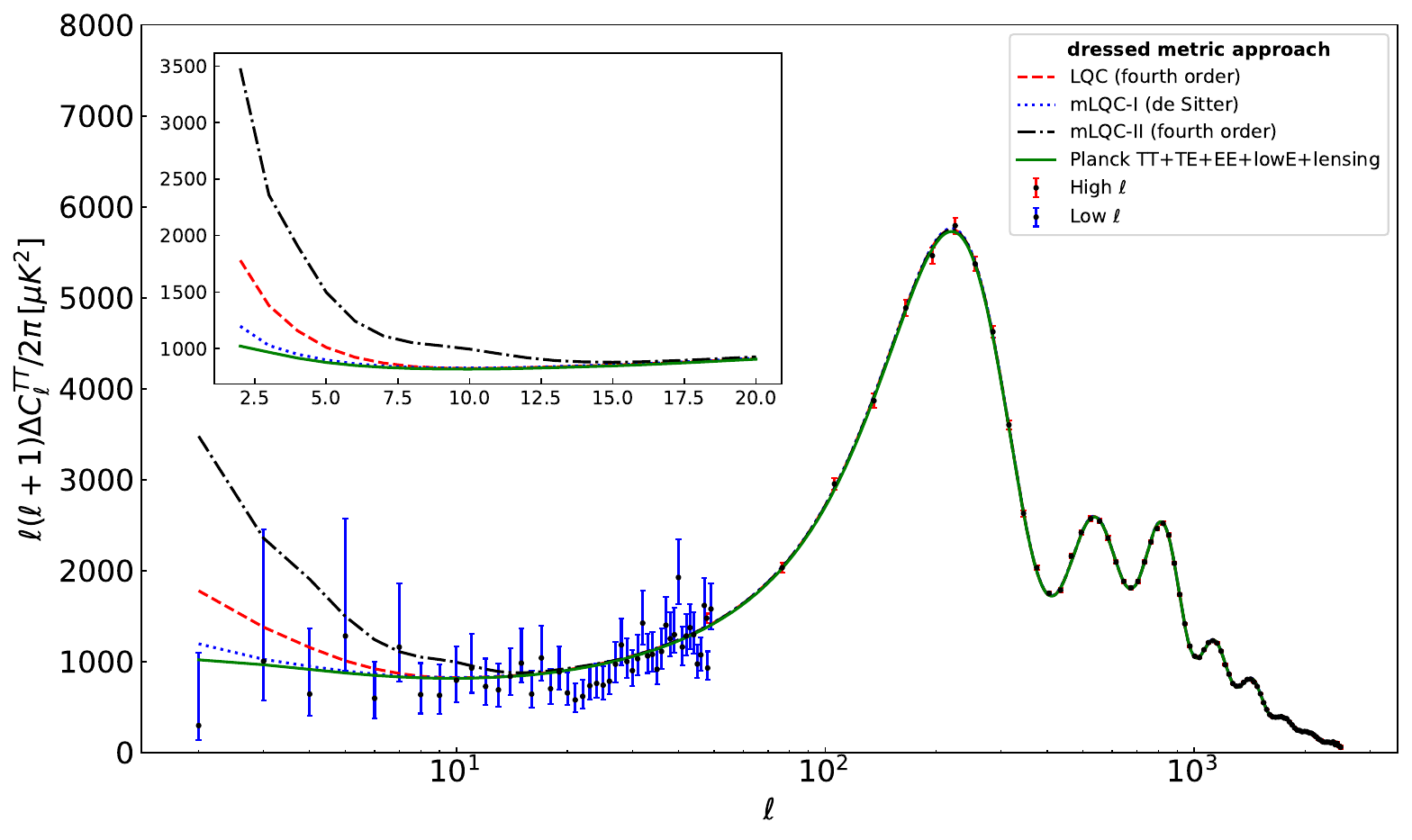}
     \caption{The angular power spectrum predicted by LQC models in the case of the dressed metric approach, while fourth order adiabatic initial states are used for LQC and mLQC-II and de Sitter initial state is used for mLQC-I. The black dots are the Planck 2018 temperature angular power spectrum, with blue error bars for low $l$ multipoles and red error bars for large $l$ multipoles. The green curve is the $\Lambda$CDM angular power spectrum best fit to Planck 2018 data.}
    \label{angular-power-fourth-order-dressed}
\end{figure}

The results for the angular power spectrum are compared in Figs. \ref{angular-power-zero-order-hybrid}-\ref{angular-power-fourth-order-dressed}. We plot the angular power spectrum for zeroth order adiabatic initial states for the hybrid approach in Fig. \ref{angular-power-zero-order-hybrid} and for the dressed metric approach in Fig. \ref{angular-power-zero-order-dressed} in all three models, namely LQC and mLQCs. However, we should point out that for mLQC-I in the dressed metric approach, we compare the angular power spectrum with the primordial power spectrum in the right panel of Fig. \ref{power-spectum-mLQC-I} which is computed for the de Sitter initial state. The black dots are the Planck 2018 temperature angular power spectrum, with blue error bars for low $l$ multipoles and red error bars for large $l$ multipoles. The green curve is the $\Lambda$CDM angular power spectrum best fit to the Planck collaboration in the TT,TE,EE-lowE+lensing 68\% limits data. From Figs. \ref{angular-power-zero-order-hybrid} and \ref{angular-power-zero-order-dressed}, it is clear that all three models match the best fit curve at large multipoles $l$ in both the hybrid approach and the dressed metric approaches, since they predict the same scale-invariant regime at large $k$. However, all three curves deviate from the best fit curve from $\Lambda$CDM model and predict a larger angular power spectrum at large angles, i.e., low $l$ multipoles due to amplification of the primordial power spectrum in the part of the intermediate regime that is next to the almost scale-invariant regime. In fact, mLQC-II exhibits the largest amplitude for the angular power spectrum at large angles, followed by LQC, and finally mLQC-I. Therefore, mLQC-I is the one that can produce the result closest to that from the $\Lambda$CDM model in both the hybrid and the dressed metric approaches. Moreover, comparing the angular power spectrum from the hybrid approach with the one from the dressed metric approach for the same model, one can find that the deviation from the best fit curve at large angles is larger in the case of the dressed metric approach. This is because the amplitude of the intermediate regime, $k \in (0.01, 10)$, is larger in the case of the dressed metric approach right next to the scale-invariant regime.

In addition, the results for the angular power spectrum for the second order adiabatic initial states in LQC and mLQCs are given in Fig. \ref{angular-power-second-order-hybrid} for the hybrid approach and in Fig. \ref{angular-power-second-order-dressed} for the dressed metric approach. Note that for mLQC-I in the dressed metric approach, similar to the case with the zeroth order adiabatic initial states, we compare the angular power spectrum obtained from the primordial power spectrum computed with the de Sitter initial state. One can find from the figures that the amplitude of the angular power spectrum from the second order adiabatic states is almost the same as that of the zeroth order adiabatic initial states. That is because although the primordial power spectrum has a smaller amplitude for the second order adiabatic initial states in the left part of the intermediate regime at very small $k$, they actually have very similar magnitude in the right part of the intermediate regime next to the almost scale-invariant regime. Based on these results, only the latter contributes mostly to the angular power spectrum at low $l$ multipoles.

In Figs. \ref{angular-power-fourth-order-hybrid} and \ref{angular-power-fourth-order-dressed}, we compare the angular power spectrum predicted by LQC and mLQCs with the fourth order adiabatic initial states in both the hybrid and the dressed metric approaches. As one can see from Fig. \ref{angular-power-fourth-order-hybrid}, LQC has a smaller angular power spectrum at large angles in comparison with mLQC-I, while the reverse is true for the zeroth and second order adiabatic initial states in the case of the hybrid approach. The reason is the presence of a large spike right before the scale-invariant regime in the primordial power spectrum with fourth order adiabatic initial states. However, the angular power spectrum for fourth order adiabatic initial states in the case of the dressed metric approach is identical to zero and second order adiabatic initial states. The reason is that the difference in the intermediate regime due to the different order of adiabatic initial states occurs in very small $k$, so the contribution to angular power is tiny and almost indistinguishable.

Finally, we summarize this section by pointing out that the predictions for the primordial power spectrum and the relevant angular power spectrum depend on the regularization ambiguities in the background dynamics, quantum ambiguities originating from treatments of cosmological perturbations, the order of adiabatic initial states, and also how far from the bounce they are imposed in the contracting branch. In fact, we realize that although the angular power spectra computed in all three models and two different perturbation approaches are consistent with the CMB observations at small scales with $l\ge20$, they actually exhibit different behaviors at large angles for low $l$ multipoles. In general, the angular power spectrum computed using the hybrid approach has a smaller deviation from the angular power spectrum predicted by the standard $\Lambda$CDM cosmological model in comparison with the dressed metric approach. Besides, among these three models, mLQC-I shows the smallest deviation from the angular power spectrum predicted by the standard $\Lambda$CDM cosmological model at large angles for zeroth and second order adiabatic initial states, while for the fourth order adiabatic initial states in the hybrid approach, LQC has the smallest deviation from the angular power spectrum predicted by the standard $\Lambda$CDM cosmological model at large angles. In any case, mLQC-II has the largest deviations from the angular power spectrum predicted by the standard $\Lambda$CDM cosmological model at large angles.

\section{Summary}
\lb{sec:conclusion}

In this manuscript, we conduct a detailed investigation on the primordial power spectrum and the relevant angular power spectrum in loop quantum cosmological models for a spatially flat FLRW universe filled with a single inflationary scalar field. Our main purpose is to investigate the potential observational signals from CMB that can be used to distinguish three loop quantum cosmological models, namely the standard LQC and Thiemann regularized versions mLQC-I/II, arising from regularization ambiguities in the background dynamics, as well as to look for signals to differentiate two perturbation approaches, namely the dressed metric approach and the hybrid approach in LQC. We first briefly reviewed the background dynamics of these three models for a spatially flat FLRW spacetime, with an emphasis on the effective dynamics of each model. Using the effective Hamiltonian constraint, one can first derive the effective Hamilton's equations in each model and then numerically solve the evolution of the background dynamics of the universe with a given set of initial conditions. To facilitate the comparison of three models, we choose the inflationary potential to be the Starobinsky potential, which is favored by Planck 2018 data, and the initial conditions are set at the bounce point with a particular choice of the value of the inflaton field so that the duration of the inflationary phase in each model is fixed to be the same number of e-foldings $N_{e}=65$. While the analysis presented in this manuscript focused on these number of e-foldings, our results did not change when the e-foldings were changed to 60 or 70. 

Once the background dynamics is fixed, we then proceed with the linear cosmological perturbations on the quantum background spacetimes. To numerically compute the primordial power spectrum and the angular power spectrum in each model, we appeal to two alternative perturbation approaches, namely the dressed metric approach and the hybrid approach, which both use Fock quantized perturbations on the loop quantized background. Our previous work has demonstrated that the difference in these two approaches at a practical level is tied to the way polymerization is performed at different steps and at a phenomenological level both the approaches are closely related \cite{Li:2022evi}. In particular, using the effective dynamics, the modified Mukhanov-Sasaki equation for each model in the dressed metric and the hybrid approaches can be obtained by polymerizing the background quantities, namely the inverse of the conjugate momentum of the scale factor and its square, in the classical Mukhanov-Sasaki equation, and this procedure leads to distinct effective mass functions for different models and approaches as their unique features when compared with one another. Equipped with the modified Mukhanov-Sasaki equation, we then move on to numerically compute the primordial power spectrum in each model and perturbation approach. For the initial states of the linear perturbations, we choose the zeroth, second, and fourth order adiabatic initial states in the contracting branch when the adiabatic conditions are satisfied, with an exception to the mLQC-I in the dressed metric approach, whose initial states are chosen to be the exact de Sitter solution tailored to the special properties of the effective mass function in this model and approach. To compare these models appropriately, we set the inflaton's mass and the initial value of the scalar field at the bounce in such a way that all models predict not only approximately the same number of inflationary e-foldings, which is $N_{e}=65$, but also almost the same scale-invariant regime for the primordial power spectrum with a relative difference of less than one percent. As a result, all the differences in the predicted primordial power spectrum in LQC and mLQC-I/II from the dressed metric approach and the hybrid approach can be traced to the differences in the infrared and the intermediate regimes, which are supposed to encode the quantum gravitational effects. 

From the resulting primordial power spectrum for each model and approach, we find some interesting results. Firstly, the moment when the adiabatic initial states of the linear perturbations are imposed in the contracting phase can affect the amplitude of the primordial power spectrum in the intermediate regime. This is true for all three models in both the dressed metric and the hybrid approach, except mLQC-I in the dressed metric approach, in which the exact de Sitter initial state is employed. The amplitude of the primordial power spectrum increases when the adiabatic initial states are chosen at an earlier time in the contracting phase. This is mainly because these states are just the approximate solutions of the Mukhanov-Sasaki equation of the mode function. In contrast, in the dressed metric approach of mLQC-I, the exact solution of the Mukhanov-Sasaki equation, which is the de Sitter initial state, is available, and the resulting primordial power spectrum is then independent of the initial time. Secondly, when different adiabatic initial states are employed, the amplitude of the primordial power spectrum also depends on the order of these states. To be specific, irrespective of the perturbation approach, in LQC and mLQC-II, the fourth order adiabatic initial states result in the primordial power spectrum with the lowest amplitude in the intermediate regime as compared to the zeroth and second order adiabatic initial states. When the order of the adiabatic initial states decreases, the amplitude of the primordial power spectrum increases. The only exception to this observation is mLQC-I in the hybrid approach, where we find the primordial power spectrum resulting from the fourth order adiabatic initial states has a larger amplitude than that from the second order adiabatic initial states. Moreover, in this case, there appears a peak of large magnitude in the rightmost part of the intermediate regime, which precedes the scale-invariant regime. Thirdly, the primordial power spectrum for both LQC and mLQC-II in the dressed metric approach has a slightly stronger suppressing regime in the infrared regime in comparison with the hybrid approach. Moreover, for mLQC-I in the dressed metric approach with de Sitter initial state, the primordial power spectrum reaches a constant value in the infrared regime rather than being suppressed. Finally, from the primordial power spectrum, it is not easy to distinguish the dressed metric approach from the hybrid approach in LQC and mLQC-II, where the amplitude of the primordial power spectrum in the intermediate regime in two approaches looks close to each other once the same initial states are chosen. The differences in primordial power spectrum between these two approaches become only discernible in mLQC-I, where one has to choose different initial states for the two approaches. As a result, to distinguish the observational effects of these two approaches and the regularization ambiguities, one must go through further steps to compute the angular power spectrum in each model and approach.

When the numerical primordial power spectrum is fed into the CAMB code, the relevant angular power spectrum for each model and approach can be obtained. From our results, we find that although all the models and approaches can result in the angular power spectrum, which is consistent with the angular power spectrum predicted by the standard $\Lambda$CDM cosmological model at small scales with $l\ge20$, they do have distinct predictions on the angular power spectrum at large angles with $l<20$. To be specific, in the dressed metric approach, the order of the adiabatic initial states in LQC and mLQC-II would not affect the amplitude of the angular power spectrum at large angles with $l< 20$. Besides, the predicted angular power spectrum at large angles always has a larger deviation from the angular power spectrum predicted by the standard $\Lambda$CDM cosmological model in mLQC-II than in LQC. This immediately makes mLQC-II less appealing as compared with LQC. Furthermore, although mLQC-I predicts a primordial power spectrum with a Planck scale infrared regime, it turns out that the resulting angular power spectrum from mLQC-I is largely improved at large angles, with the smallest deviations from the angular power spectrum predicted by the standard $\Lambda$CDM cosmological model among all three models. This implies that in the dressed metric approach, the averaged amplitude of the resulting primordial power spectrum takes the lowest values in the rightmost part of the intermediate regime neighboring the scale-invariant regime since only this part of the primordial power spectrum significantly contributes to the angular power spectrum at large angles. On the other hand, in the hybrid approach, we observe a similar pattern for the zeroth and second order adiabatic initial states, while the deviation of the angular power spectrum at large angles from the angular power predicted by the standard $\Lambda$CDM cosmological model is smaller in comparison with the dressed metric approach. The deviation of the angular power spectrum at large angles from the angular power spectrum predicted by the standard $\Lambda$CDM cosmological model is always largest in mQLC-II and smallest in mLQC-I. In particular, it is worth emphasizing that with the simplest zeroth and second order adiabatic initial states set in the contracting phase, one can obtain an angular power spectrum that is close to the angular power spectrum predicted by the standard $\Lambda$CDM cosmological model even at low $l$ multipoles from mLQC-I by using the hybrid approach. As compared with the results in LQC and mLQC-II,  similar results between mLQC-I and $\Lambda$CDM are very striking since no special choice of the initial states of the linear perturbations is required. In this sense, mLQC-I seems to be a more favorable construction of the quantum cosmological theory from LQG, and the hybrid approach also appears easier to reconcile with the observations. Finally, with the fourth order adiabatic initial states, LQC results in an angular power spectrum with the least deviation from the angular power spectrum predicted by the standard $\Lambda$CDM cosmological model at large angles. In this case, the results from mLQC-I are less satisfactory due to the spike in the intermediate regime. These results are very interesting since by construction mLQC-I follows the procedure in LQG more directly than any other considered model. 

To conclude, our investigations on the angular power spectrum predicted by LQC and mLQCs models in both the dressed metric approach and the hybrid approach reveal that quantization regularization and quantum ambiguities are not merely theoretical artifacts. Instead, they can lead to potential signals that can in principle be compared and tested by direct observational data in the future. Although with the commonly used adiabatic initial states, none of the models and approaches actually resolve the anomalies in the angular power spectrum at large angular scales \cite{Planck:2018jri}, the similarities between the angular power spectrum predicted by mLQC-I and the standard $\Lambda$CDM cosmological model point out a possible new direction to resolve this issue. In particular, there might be a certain regularization that can lead to a quantum cosmological model in which the angular power spectrum is naturally suppressed by setting general initial states in the contracting phase. It will be interesting to examine these models with special initial states, as considered earlier for standard LQC, to explore whether modified versions of LQC can result in alleviation of anomalies in CMB. 

\section*{Acknowledgments}
 
B.-F. Li is supported by the National Natural Science Foundation of China (NNSFC) with the grant No. 12005186. M.M and P.S. are supported by NSF grant PHY-2110207 and PHY-2409543.

%\bibliography{BibNVar}
%\bibliographystyle{h-physrev5}

\end{document}